\documentclass[12pt]{article}

\def\bSig\mathbf{\Sigma}

\usepackage{amsmath,amssymb}
\usepackage[dvips]{graphicx}
\usepackage{url,natbib,color,verbatim,xfrac,enumitem}
\usepackage{float}
\usepackage{bm,graphicx,fancyhdr}
\usepackage{caption,multirow,subfigure}
\usepackage{setspace}
\usepackage{listings}
\usepackage{JASA_manu}

\usepackage[boldmath]{numprint}

\bibpunct{(}{)}{;}{a}{}{,}

\renewcommand\emph{\textit}

\renewcommand\hat{\widehat}

\newsavebox\CBox
\def\textBF#1{{\fontseries{b}\selectfont{#1}}}

\begin{document}

{\center{\large{Generalized G-estimation and Model Selection}}}

{\center{\large{Michael P. Wallace$^1$, Erica E. M. Moodie$^2$, and David A. Stephens$^3$}}}

\vspace{0.1in}

\noindent $^1$ Department of Statistics and Actuarial Science, University of Waterloo

\noindent $^2$ Department of Epidemiology, Biostatistics, and Occupation Health, McGill University

\noindent $^3$ Department of Mathematics and Statistics, McGill University

\hrulefill

\begin{abstract}
Dynamic treatment regimes (DTRs) aim to formalize personalized medicine by tailoring treatment decisions to individual patient characteristics. G-estimation for DTR identification targets the parameters of a structural nested mean model known as the blip function from which the optimal DTR is derived. Despite considerable work deriving such estimation methods, there has been little focus on extending G-estimation to the case of non-additive effects, non-continuous outcomes or on model selection. We demonstrate how G-estimation can be more widely applied through the use of iteratively-reweighted least squares procedures, and illustrate this for log-linear models. We then derive a quasi-likelihood function for G-estimation within the DTR framework, and show how it can be used to form an information criterion for blip model selection. These developments are demonstrated through application to a variety of simulation studies as well as data from the Sequenced Treatment Alternatives to Relieve Depression study. 

\vspace{0.1in}

\end{abstract}
\newpage

\section{Introduction}

Dynamic treatment regimes (DTRs) - sequences of decision rules that take patient information as input and output recommended treatments - are part of a rapidly expanding literature on personalized medicine \citep{MOODIEBOOK}. By tailoring treatments to individual patient characteristics, DTRs are able to improve long-term outcomes for a population when compared with more traditional non-tailored approaches. Identification of the \emph{optimal} regime (which maximizes expected outcome) is a major challenge due to, for example, delayed treatment effects and covariate-dependent treatment assignment.

Numerous methods have been proposed for optimal DTR estimation.  A general class of DTR estimation approaches relies on structural nested mean models (SNMMs, \citealt{ROBINS1994,VANSTEELSNM2003}).  In our formulation, the SNMM parameterizes the difference between the conditional expectation of the outcome following observed treatment with that of a counterfactual outcome under a (potentially unobserved) treatment regime. By estimating the parameters of this model we are then able to identify the optimal DTR, i.e.~the sequence of treatment decisions that maximizes the expected outcome across all patients. This general approach of parameterizing and estimating components of the outcome mean model is used in a variety of specific DTR estimation methods, including Q-learning \citep{WATKINS1989REINFORCE,SUTTON1998REINFORCE}, dynamic weighted least squares \citep{WallacedWOLS}, and G-estimation \citep{ROBINS2004PROC}, the last of which is the focus of this paper.

Almost all of the methodological developments for DTR estimation have focused on continuous outcomes and additive effects of treatment on the expected counterfactual outcome, with time-to-event outcomes included as a special case.  Estimation for discrete outcomes, or for effects of treatment on non-additive scales has received little attention.  A recent exception is the work of \cite{Moodie2014Qlearn} who used generalized additive models to apply Q-learning in this setting. \cite{THALL2000CANCER}, meanwhile, considered a likelihood-based approach in the case of a binary outcome. Such examples are rare, however, and typically grounded in methods that do not offer a great deal of flexibility or robustness in modeling. The presentation (and implementation) of G-estimation primarily for continuous outcomes therefore represents an important limitation of the approach.

Even for continuous outcomes, there has been little focus on model selection in the context of G-estimation. The methods listed above all implicitly assume that the SNMMs upon which they rely are correctly (or possibly over-) specified. Very little work has been published related to the problem of choosing between a set of candidate models or model checking. Exceptions include the diagnostic plots of \cite{RICH2010AZT} and the method of \cite{Wallace2016DRAss} that exploits the so-called double-robustness property (discussed below) for model assessment. Neither of these, however, assess the component of the model quantifying the effect of treatment -- i.e., the \textit{blip} model -- alone, and can at best assess the validity of both the blip model \emph{and} another component model simultaneously.


In this paper, we present two generalizations of G-estimation. First, we derive and illustrate how iteratively-reweighted least squares (IRLS) may be used to implement G-estimation in a discrete-outcome scenario using log-linear models.  We then present a new approach to model selection when using G-estimation for DTRs based on a Quasi-likelihood Information Criterion (QIC). Our QIC formulation is applicable to G-estimation procedures in general, but 
we demonstrate how the QIC can be applied to G-estimation in the DTR setting, encompassing multiple stages of treatment 
for both the cases of continuous and count outcomes.

\section{DTRs and G-estimation}
\label{sec:DTRIntro}

We establish notation by considering G-estimation in its conventional form, where effects of exposure are additive and a linear model is presumed for the counterfactual outcomes.  We consider a cohort of subjects on whom data are gathered at fixed intervals (such as visits to a physician) or at fixed clinical decision points (diagnosis, remission, and so on), with a treatment decision made at each of these time points. Our objective is to identify the sequence of treatment decision rules (the DTR) which maximizes a subject's long-term expected outcome (defined such that larger values are preferred). We assume there are a total of $J$ successive treatment decisions (or \emph{stages}): $y$ denotes observed patient outcome; $a_j$ denotes the stage $j$ treatment decision ($j = 1, ..., J$), with $a_j^0$ denoting ``no treatment'' (such as a control or standard care); $\bm{h}_j$ denotes the covariate matrix containing patient information (\emph{history}) prior to the $j^{th}$ treatment decision.  The history can include previous treatments $a_1,...,a_{j-1}$ along with non-treatment information $\bm{x}_j$.  In addition, over- and underline notation is used to indicate the past and future, respectively. For example $\overline{\bm{a}}_j$ denotes the vector of treatment decisions up to and including the stage $j$ decision, while $\underline{\bm{a}}_{j+1}$ denotes the last $J-j$ decisions (from stage $j+1$ up to and including stage $J$). The \emph{optimal} treatment at any given stage is denoted ${a}_j^{opt}$.

The (stage $j$) \emph{optimal blip-to-reference} (or simply \emph{blip}) function is defined as
\begin{eqnarray*}
{\gamma}_j(\bm{h}_j,{a}_j) = E[{Y}(\overline{\bm{a}}_{j-1},a_j,\underline{\bm{a}}_{j+1}^{opt}) - {Y}(\overline{\bm{a}}_{j-1},{a}_j^0,\underline{\bm{a}}_{j+1}^{opt})| \bm{h}_j]
\end{eqnarray*}

\noindent which is the expected difference in outcome when using a reference treatment ${a}_j^0$ instead of ${a}_j$ at stage $j$, in subjects with history $\bm{h}_j$ who receive optimal treatment across the remaining $J-j$ intervals ($\underline{\bm{a}}_{j+1}^{opt}$). The optimal treatment at stage $j$ maximizes the blip. Under additive local rank preservation (see 2.1.3 of \citealt{MOODIEBOOK}), we can decompose the expectation of the observed potential outcome as
\begin{eqnarray*}
E[{Y(\underline{\bm{a}}_J)}] = E[{Y}^{opt}] - \sum_{j=1}^J \left[{\gamma}_j(\bm{h}_j,{a}_j^{opt}) - {\gamma}_j(\bm{h}_j,{a}_j) \right]
\end{eqnarray*}

\noindent where ${Y}^{opt}$ can be thought of as the optimal outcome that would be observed if the optimal treatment was followed at every stage. The observed outcome ${y}$ is then equal in expectation to the optimal outcome minus the difference in outcome between optimal and observed treatment at each stage.

In practice, we assume $\gamma_j(.)$ takes a known parametric form ${\gamma}_j(\bm{h}_j,{a}_j;\bm{\psi}_j)$ with parameters $\bm{\psi}_j$.  We then estimate $\bm{\psi}_j$, and identify the optimal treatment regime by choosing, for each subject, the treatment that maximizes the estimated blip.  G-estimation is one method which may be used to estimate $\bm{\psi}_j$, and relies on two standard assumptions: the \emph{stable unit treatment value assumption} and the assumption of \emph{no unmeasured confounding} (or \emph{sequential randomization}). The former means that a subject's outcome is not influenced by other subjects' treatment allocation \citep{RUBIN1980SUTVA} and that the counterfactual outcome under a particular treatment is equal to the observed outcome under that treatment; the latter states that the treatment received at stage $j$ is independent of any future (potential) covariate or outcome, conditional on history $\bm{h}_j$.

Writing $\bm{\underline{\psi}}_j = (\bm{\psi}_j,\bm{\psi}_{j+1},...,\bm{\psi}_{J})$, we define for each $j$, ${G}_j(\bm{\underline{\psi}}_j) = \widetilde{{y}}_j - {\gamma}_j(\bm{h}_{\psi j},{a}_j;\bm{\psi}_j)$  where $\widetilde{{y}}_j = {y} + \sum_{k=j+1}^J \left[{\gamma}_k(\bm{h}_{\psi k},{a}_k^{opt};\bm{\psi}_k) - {\gamma}_k(\bm{h}_{\psi k},{a}_k;\bm{\psi}_k) \right]$ can be viewed as a \emph{pseudo-outcome} which we compute at each stage based on those $\hat{\bm{\psi}}_k$ ($k > j$), and hence $\hat{a}_k^{opt}$ already estimated. Therefore
\begin{eqnarray*}
{G}_j(\bm{\underline{\psi}}_j) &=& {y} - {\gamma}_j(\bm{h}_j,{a}_j;\bm{\psi}_j) + \sum_{k=j+1}^J \left[{\gamma}_k(\bm{h}_k,{a}_k^{opt};\bm{\psi}_k) - {\gamma}_k(\bm{h}_k,{a}_k;\bm{\psi}_k) \right]
\end{eqnarray*}
\noindent

\noindent and we can regard ${G}_j(\bm{\underline{\psi}}_j)$ as being equal to the expected outcome with the effects of stage $j$ treatment `removed' and the difference between optimal and observed treatment thereafter `added'. Under the above assumptions we have that $E[{G}_j(\bm{\underline{\psi}}_j)|\bm{h}_j] = E[Y(\overline{\bm{a}}_{j-1},{a}_j^0,\underline{\bm{a}}_{j+1}^{opt})|\bm{h}_j]$ which represents the expected outcome for a subject who receives treatment history $\overline{\bm{a}}_{j-1}$ up to stage $j-1$, \textit{no} treatment at stage $j$, and optimal treatment thereafter. We refer to ${G}_j$ as the stage $j$ \emph{treatment-free} outcome.

To estimate the blip parameters $\bm{\psi}_j$, G-estimation considers the set of functions
\begin{eqnarray*}
{U}_j(\bm{\underline{\psi}}_j;\bm{\beta}_j;\bm{\alpha}_j) = \left\{{S}_j({A}_j) - E[{S}_j({A}_j)|\bm{h}_j;\bm{\alpha}_j]\right\}\{{G}_j (\bm{\underline{\psi}}_j) - E[{G}_j (\bm{\underline{\psi}}_j)|\bm{h}_j;\bm{\beta}_j]\}
\end{eqnarray*}
\noindent where typically ${S}_j({A}_j) = {a}_j\bm{h}_j$. A fully efficient form of ${S}_j({A}_j)$ has been proposed \citep{ROBINS2004PROC}, but requires knowledge of the variance of ${G}_j(\bm{\underline{\psi}}_j)$, which is rarely available in practice.  The estimating functions require the specification of a number of models, namely the stage $j$ blip model: ${\gamma}_j(\bm{h}_{\psi j},{a}_j;\bm{\psi}_j)$; the stage $j$ treatment-free model: $E[{G}_j (\bm{\underline{\psi}}_j)|\bm{h}_{\beta j};\bm{\beta}_j]$; and the stage $j$ \emph{treatment} model: $E[{A}_j|\bm{h}_{\alpha j};\bm{\alpha}_j]$ (or, more generally, $E[{S}_j({A}_j)|\bm{h}_j;\bm{\alpha}_j]$), where $\bm{h}_{\psi j}$, $\bm{h}_{\beta j}$ and $\bm{h}_{\alpha j}$ are subsets of patient history that feature in the blip, treatment-free, and treatment models, respectively. An important property of G-estimation is its double-robustness: if the blip is correctly specified, then as long as at least one of the treatment and treatment-free models is also correctly specified the resulting blip parameter estimators will be consistent.

Because the functions ${G}_j(\bm{\underline{\psi}}_j)$ depend on the observed outcome ${y}$ and each blip model from stage $j$ onwards, G-estimation proceeds recursively, starting at the final stage $J$ and working backwards to stage 1. At each stage the above models are specified, and then the following three steps are carried out:
\begin{enumerate}
\item Estimate the treatment model parameters $\hat{\bm{\alpha}}_j$ by regressing the stage $j$ treatment ${a}_j$ on the treatment model covariates $\bm{h}_{\alpha j}$.
\item Estimate the treatment-free model parameters $\bm{\beta}_j$ by `regressing' ${G}_j(\bm{\underline{\psi}}_j)$ on $\bm{h}_{\beta j}$, where rather than conducting a standard least squares regression, we instead solve the corresponding least squares equation to give $\hat{\bm{\beta}}_j(\bm{\psi}_j,\bm{\underline{\hat{\psi}}}_{j+1})$ in terms of the stage $j$ blip parameters $\bm{\psi}_j$ and the estimated blip parameters $(\bm{\underline{\hat{\psi}}}_{j+1})$ from previous stages.
\item Using the estimates $\hat{\bm{\alpha}}_j$ and $\hat{\bm{\beta}}_j(\bm{\psi}_j,\bm{\underline{\hat{\psi}}}_{j+1})$ from steps 1 and 2, solve the equation $E_n[{U}_j(\bm{\psi}_j;\hat{\bm{\beta}}_j,\hat{\bm{\alpha}}_j)] = 0$ to estimate $\bm{\psi}_j$, where $E_n$ denotes the mean over all subjects.
\end{enumerate}

\noindent We can then use the resulting blip parameter estimates $\hat{\bm{\psi}}_j$ to estimate the optimal stage $j$ treatment ${a}_j^{opt}$ for each subject, and hence the function ${G}_{j-1}(\bm{\underline{\psi}}_{j-1})$, and repeat the above steps until estimates are obtained for every stage of the analysis.

\section{G-estimation for generalized linear models}

The framework in section \ref{sec:DTRIntro} is standard for G-estimation applications.  It assumes a continuous outcome, and that the treatment modifies the expected outcomes additively, that is, the blip acts additively on the original outcome scale.  The construction can be modified to be applicable to discrete outcomes, but relaxing an assumption of additivity of the treatment effect needs more care.  In this section, we demonstrate how G-estimation may be generalized to handle other effect types, and show how estimation can be achieved using standard computational approaches. Specifically, we will apply G-estimation for generalized linear models by using iteratively-reweighted least squares.

\subsection{G-estimation for multiplicative effects}

\label{sec:Multiplicative}
For an arbitrary counterfactual outcome $Y(a)$, the effect of exposure may be framed in terms of the average potential outcome $E[Y(a)]$, and contrasts comparing this average for different exposures.  For example, for a binary exposure we might consider the ratio of expectations $E[Y(1)]/E[Y(0)]$ rather than the expected ratio $E[Y(1)/Y(0)]$; this focuses on population- rather than individual-level contrasts and avoids identifiability issues associated with attempting to specify a joint model for $\{Y(0),Y(1)\}$.

For G-estimation, consider for illustration the two interval case; our approach will focus on constructing models for $E[Y(a_1,a_2)]$ using the decomposition
\[
E[Y(a_1,a_2)] = E[Y(a_1^{opt},a_2^{opt})] \frac{E[Y(a_1,a_2^{opt})|\bm{h}_1]}{E[Y(a_1^{opt},a_2^{opt})|\bm{h}_1]} \frac{E[Y(a_1,a_2)|\bm{h}_2]}{E[Y(a_1,a_2^{opt})|\bm{h}_2]}
\]
that is, using a multiplicative modification of the optimal outcome, and making a multiplicative rank preserving assumption.  In this context, the blip function ${\gamma}_j(\bm{h}_{\psi j},{a}_j)$ may be defined as the ratio of expected counterfactual outcomes $E[{Y}(\overline{\bm{a}}_{j-1},a_j,\underline{\bm{a}}_{j+1}^{opt}) | \bm{h}_{\psi j}]/E[{Y}(\overline{\bm{a}}_{j-1},{a}_j^0,\underline{\bm{a}}_{j+1}^{opt})|\bm{h}_{\psi j}]$
and the expected counterfactual outcome may be computed as
\begin{eqnarray*}
E[Y(\underline{\bm{a}}_J)|\underline{\bm{h}}_J] = E[{Y}^{opt}] \prod\limits_{j=1}^J \left[{\gamma}_j(\bm{h}_{\psi j},{a}_j)/{\gamma}_j(\bm{h}_{\psi j},{a}_j^{opt})\right]
\end{eqnarray*}
or equivalently
\begin{eqnarray*}
\log(E[{Y(\underline{\bm{a}}_J)}|\underline{\bm{h}}_J]) = \log(E[{Y^{opt}}])  - \sum_{j=1}^J \left[\log\left({\gamma}_j(\bm{h}_{\psi j},{a}_j^{opt})/{\gamma}_j(\bm{h}_{\psi j},{a}_j)\right) \right],
\end{eqnarray*}
giving rise to a stage-$j$ pseudo-outcome analogous to that in the continuous outcome, linear model setting as
\begin{equation}\label{eq:ytildeloglinear}
\widetilde{{y}}_j = {y} \times \prod_{k=j+1}^J \left[\left({\gamma}_k(\bm{h}_{\psi k},{a}_k^{opt};\bm{\psi}_k)/{\gamma}_k(\bm{h}_{\psi k},{a}_k;\bm{\psi}_k)\right) \right],
\end{equation}
and hence ${G}_j(\bm{\underline{\psi}}_j) = \widetilde{{y}}_j/{\gamma}_j(\bm{h}_{\psi j},{a}_j;\bm{\psi}_j)$. We then propose log-linear models for the treatment-free and blip models
\begin{eqnarray*}
\log\left\{E[{G}_j(\bm{\underline{\psi}}_j)|\bm{h}_{\beta j};\bm{\beta}_j]\right\} = \bm{h}_{\beta j} \bm{\beta}_j \qquad \qquad
\log \gamma_j(\bm{h}_{\psi j},{a}_j;\bm{\psi}_j) = {a}_j\bm{h}_{\psi j} \bm{\psi}_j,
\end{eqnarray*}
from which, via some rearrangement, the G-estimating functions become
\begin{eqnarray*}
{U}_j(\bm{\underline{\psi}}_j;\bm{\beta}_j;\bm{\alpha}_j) &=& \left\{{S}_j({A}_j) - E[{S}_j({A}_j)|\bm{h}_j;\bm{\alpha}_j]\right\}\{{G}_j (\bm{\underline{\psi}}_j) - E[{G}_j (\bm{\underline{\psi}}_j)|\bm{h}_{\beta j};\bm{\beta}_j]\}\\
&=& \left\{{a}_j - E[{A}_j|\bm{h}_{\alpha j};\bm{\alpha}_j]\right\}\{\widetilde{{y}}_j/\exp({a}_j\bm{h}_{\psi j} \bm{\psi}_j) - \exp(\bm{h}_{\beta j} \bm{\beta}_j)\} \bm{h}_{\psi j}.
\end{eqnarray*}
Again suppressing stage-specific notation, and introducing subscript-$i$ notation for subject $i$, G-estimation at each stage thus solves
\begin{eqnarray}\label{eq:poiseq1}
0 = \sum_{i=1}^n d_i \bm{h}_{\psi i} (\widetilde{{y}}_i - \mu_i(\bm{\beta},\bm{\psi})),
\end{eqnarray}
with $d_i = a_i - E[{A}_i|\bm{h}_{\alpha i};\bm{\alpha}]$ and $\mu_i(\bm{\beta},\bm{\psi}) = \exp(\bm{h}_{\beta i} \bm{\beta} + {a}_i \bm{h}_{\psi i} \bm{\psi})$. In section \ref{sec:IRLS}, we present an IRLS algorithm to estimate blip parameters at each stage in the usual recursive manner.

Note that when the observed $y$ value is zero, the pseudo-outcomes in \eqref{eq:ytildeloglinear} will also be zero unless a further adjustment is made. A simple approach to this issue is to assume that when $y=0$, it is drawn from a Poisson distribution with mean 0.001 (or some other small value), and replace $y$ with its expectation. For example, for stage $J-1$, after parameters $\bm{\psi}_J$ are estimated, the usual adjustment $\widetilde{{y}}_{J-1} = y \times \left({\gamma}_J(\bm{h}_{\psi J},\widehat {a}_J^{opt};\widehat{\bm{\psi}}_J)/{\gamma}_J(\bm{h}_{\psi J},{a}_J;\widehat{\bm{\psi}}_J)\right)$
becomes
\[
\widetilde{{y}}_{J-1} = 0.001 \times \left({\gamma}_J(\bm{h}_{\psi J},\widehat {a}_J^{opt};\widehat{\bm{\psi}}_J)/{\gamma}_J(\bm{h}_{\psi J},{a}_J;\widehat{\bm{\psi}}_J)\right)
\]
if $y$ is zero; $\widetilde{{y}}_{J-1}$ is guaranteed non-negative.

\subsection{Iteratively-reweighted least squares}

\label{sec:IRLS}


We now demonstrate how G-estimation may proceed for log-linear models using IRLS. Without loss of generality we consider a single-stage example allowing us to suppress stage-specific notation. The G-estimation equations, as written in (\ref{eq:poiseq1}) are of a standard form from which IRLS may be used to estimate $\bm{\psi}$.  Suppose ${y}$ has a mean function $\mu$ modeled using link function $g(\cdot)$ and linear predictor vector ${\eta} = \bm{h}_\beta \bm{\beta} + {a} \bm{h}_\psi \bm{\psi}$ such that ${\mu} = g^{-1}({\eta})$.  Denote the variance function $V(\mu)$. We can then estimate $\bm{\psi}$ via IRLS using the following algorithm:

\noindent 1. Set initial parameters $\widehat{\bm{\beta}}^{(0)}$, $\widehat{\bm{\psi}}^{(0)}$ and compute for each subject the initial linear predictor $\widehat{{\eta}}^{(0)}= \bm{h}_\beta \widehat{\bm{\beta}}^{(0)} + a \bm{h}_\psi \widehat{\bm{\psi}}^{(0)}$ and mean value $\widehat{{\mu}}^{(0)} = g^{-1}(\widehat{{\eta}}^{(0)})$.

\noindent 2. Set $\widehat{z}_{i}^{(1)}=\widehat{\eta}_{i}^{(0)}+ ( y_{i}-\widehat{\mu}_{i}^{(0)} ) \dot{g} ( \widehat{\mu}_{i}^{(0)})$, $\widehat{w}_{i}^{(1)} = w_{i}/[\{ \dot{g} ( \widehat{\mu }_{i}^{(0)} )\}^{2} V ( \widehat{\mu }_{i}^{(0)})]$. Denote by $\bm{D}^{(1)}$ the diagonal matrix with $(i,i)$ element $\widehat{d}_{i}^{(1)}$, and by $\bm{A}$ the diagonal matrix with $(i,i)$ element $a_{i}$, the observed treatment for subject $i$.

\noindent 3. Apply the G-estimation procedure to re-estimate $\bm{\beta}$ and $\bm{\psi}$:
\begin{eqnarray*}
\widehat{\bm{\psi}}^{(1)} & =& \left[\bm{h}_\psi^\top (\mathbf{I}_n - \bm{h}_{\beta D})\bm{D}^{(1)} \bm{A} \bm{h}_\psi\right]^{-1} \left[\bm{h}_\psi^\top (\mathbf{I}_n - \bm{h}_{\beta D})\bm{D}^{(1)} \widehat{\bm{z}}^{(1)}\right], \\[6pt]
\widehat{\bm{\beta}}^{(1)}  &=& (\bm{h}_\beta^\top \bm{D}^{(1)} \bm{h}_\beta)^{-1}\bm{h}_\beta^\top \bm{D}^{(1)}(\widehat{\bm{z}}^{(1)} - \bm{A} \bm{h}_\psi \widehat{\bm{\psi}}^{(1)})
\end{eqnarray*}
where $\bm{h}_{\beta D} = \bm{h}_\beta (\bm{h}_\beta^\top \bm{D}^{(1)} \bm{h}_\beta)^{-1} \bm{h}_\beta^\top \bm{D}^{(1)}$, and $\mathbf{I}_n$ is the $n \times n$ identity matrix.

\noindent 4. Define vectors $\widehat{\bm{\eta}}^{(1)}=\bm{h}_\beta \widehat{\bm{\beta}}^{(1)} + \bm{A} \bm{h}_\psi \widehat{\bm{\psi}}^{(1)}$ and $\widehat{\bm{\mu}} ^{(1)}=g^{-1}\left( \widehat{\bm{\eta}}^{(1)}\right)$.

\noindent 5.  Return to 2 and iterate through 2-5 using $\widehat{\bm{\mu}}^{(1)}$ and $%
\widehat{\bm{\eta}}^{(1)}$ as the updated starting values, obtaining ($%
\widehat{\bm{\mu}}^{(2)}$,$\widehat{\bm{\eta}}^{(2)}$), then repeat to generate ($%
\widehat{\bm{\mu}}^{(3)}$,$\widehat{\bm{\eta}}^{(3)}$), and so on.

\noindent 6.  Repeat until $\widehat{\bm{\mu}}^{(t)}$ and $\widehat{\bm{\eta}}^{(t)}$
satisfy $\left| \widehat{\bm{\mu}}^{(t)}-\widehat{\bm{\mu}}^{(t-1)}\right| <\epsilon _{\mu
}$ and/or $\left| \widehat{\bm{\eta}}^{(t)}-\widehat{\bm{\eta}}^{(t-1)}\right|
<\epsilon _{\eta }$ for tolerances $\epsilon _{\mu }$ and $\epsilon _{\eta }.$

\medskip

\noindent
The sequence of estimates produced by this algorithm converges to the solution of the G-estimating equations.

\section{G-estimation and quasi-likelihood}

Inference for SNMMs using G-estimation, unlike inference for more conventional models such as generalized linear models (GLMs), is not likelihood-based and so established model selection approaches such as Akaike's Information Criterion (AIC, \citealt{Akaike1973AIC}) cannot be directly used within the DTR framework. However, we shall reframe the preceding presentations of G-estimation to illustrate how quasi-likelihood theory may be applied.

\subsection{Linear case}  We assume the treatment-free model is linear in $\bm{h}_{\beta j}$, i.e. that $E[{G}_j (\bm{\underline{\psi}}_j)|\bm{h}_{\beta j};\bm{\beta}_j] = \bm{h}_{\beta j} \bm{\beta}_j$. Then by ordinary least squares, we may estimate $\bm{\beta}_j$ as
\begin{eqnarray*}
\hat{\bm{\beta}}_j(\bm{\psi}_j,\bm{\underline{\hat{\psi}}}_{j+1}) = \left[\bm{h}_{\beta j}^\top \bm{h}_{\beta j}\right]^{-1} \bm{h}_{\beta j}^\top(\widetilde{\bm{y}}_j - \bm{A}_j\bm{h}_{\psi j} \bm{\psi}_j).
\end{eqnarray*}
For convenience we again suppress the subscript-$j$ notation -- all of what follows may be applied on a stage-by-stage basis - and write $\hat{\bm{h}}_\beta = \bm{h}_{\beta j} \left[\bm{h}_{\beta j}^\top \bm{h}_{\beta j}\right]^{-1} \bm{h}_{\beta j}^\top$. Substituting the estimate of $\hat{\bm{\beta}}$ in terms of $\bm{\psi}$, we may rewrite the estimating function vector as
\begin{eqnarray*}
{U}(\bm{\psi}) = (\bm{D}\bm{h}_\psi)^\top(\widetilde{\bm{y}} - \bm{h}_\beta \hat{\bm{\beta}}(\bm{\psi}) - \bm{A} \bm{h}_\psi \bm{\psi}) &=& (\bm{D}\bm{h}_\psi)^\top(\widetilde{\bm{y}} - \hat{\bm{h}}_\beta (\widetilde{\bm{y}} - \bm{A} \bm{h}_\psi \bm{\psi}) - \bm{A} \bm{h}_\psi \bm{\psi})\\
&=& (\bm{D}\bm{h}_\psi)^\top \left[(\mathbf{I}_n - \hat{\bm{h}}_\beta) (\widetilde{\bm{y}} - \bm{A} \bm{h}_\psi \bm{\psi})\right]\\
&=& \bm{h}_\psi^\top \bm{W} (\widetilde{\bm{y}} - \bm{A} \bm{h}_\psi \bm{\psi})
\end{eqnarray*}

\noindent where $\bm{W} = \bm{D}^\top (\mathbf{I}_n - \hat{\bm{h}}_\beta)$. From here, the estimation of $\bm{\psi}$ follows by
\begin{eqnarray}\label{eq:gestred}
\hat{\bm{\psi}} = \left(\bm{h}_\psi^\top \bm{W} \bm{A} \bm{h}_\psi\right)^{-1} \bm{h}_\psi^\top \bm{W} \widetilde{\bm{y}}.
\end{eqnarray}
The form of this estimator is straightforward (and is almost identical to a standard weighted ordinary least squares estimator). This affords greater simplicity in implementation, as well as giving a clear indication that quasi-likelihood methods may be easily applied.

We follow \citet{WedderbernQL1974} in defining the quasi-likelihood of $\bm{\psi}$ by writing $\bm{\mu} = \bm{A}\bm{h}_\psi \bm{\psi}$ and solving
\begin{eqnarray*}
\frac{\partial Q}{\partial \bm{\psi}} = \frac{\partial Q}{\partial \bm{\mu}} \frac{\partial \bm{\mu}}{\partial \bm{\psi}} = \bm{h}_\psi^\top \bm{W} (\widetilde{\bm{y}} - \bm{\mu}),
\end{eqnarray*}
yielding
\begin{eqnarray} \label{eq:QIClinear}
Q(\bm{\psi}) &=& \bm{\psi}^\top \bm{h}_\psi^\top \bm{W} \widetilde{\bm{y}} - \frac{1}{2}\bm{\psi}^\top \bm{h}_\psi^\top \bm{D} \bm{A} \bm{h}_\psi \bm{\psi} = \bm{\psi}^\top \bm{m} - \frac{1}{2}\bm{\psi}^\top \bm{M} \bm{\psi}
\end{eqnarray}
\noindent where $\bm{m} = \bm{h}_\psi^\top \bm{W} \widetilde{\bm{y}}$, $\bm{M} = \bm{h}_\psi^\top \bm{W} \bm{A} \bm{h}_\psi = \bm{h}_\psi^\top \bm{D} (\mathbf{I}_n - \hat{\bm{h}}_\beta) \bm{A} \bm{h}_\psi$, and we ignore the constant term. Because $\mathbf{I}_n - \hat{\bm{h}}_\beta$ is positive definite, $\bm{M}$ is positive semi-definite in expectation, and thus provided $n$ is large, this quasi-likelihood is uniquely maximized at $\hat{\bm{\psi}}$ in large samples. Furthermore, given the stage-by-stage, recursive nature of the G-estimation approach within the DTR setting, we may derive this quasi-likelihood at each stage of an analysis.

\subsection{Log-linear case}

In the log-linear case we first reformulate (\ref{eq:poiseq1}), dividing through by $\exp(\bm{h}_\beta \bm{\beta})$ to give
\begin{eqnarray}\label{eq:poiseq2a}
0 = \sum_{i=1}^n d_i \bm{h}_{\psi i} ({y}_i^* - \mu_i^*(\bm{\psi}))
\end{eqnarray}
where ${y}_i^* = {y}_i \exp(-\bm{h}_{\beta i} \bm{\beta})$ and $\mu_i^*(\bm{\psi}) = \exp(a_i \bm{h}_{\psi i} \bm{\psi})$. This moves the nuisance parameters $\bm{\beta}$ into a pseudo-outcome $y^*$, framing the estimating equations more explicitly in terms of the target blip parameters $\bm{\psi}$, as in the linear case. This allows us to return to the theory of Wedderburn and proceed as before by solving
\begin{eqnarray}\label{eq:loglinearG}
\frac{\partial Q}{\partial \bm{\psi}} = \frac{\partial Q}{\partial \bm{\mu^*}} \frac{\partial \bm{\mu^*}}{\partial \bm{\psi}} = \bm{h}_\psi^\top \bm{D} (\widetilde {\bm{y}}^* - \bm{\mu}^*),
\end{eqnarray}
but this does not yield a quasi-likelihood in a simple way. However, by appealing to the IRLS procedure, and a recursive calculation, we compute a quasi-likelihood suitable for model comparison by considering the sequence of linear approximations to the log-linear estimating equations implied by \eqref{eq:QIClinear}.  The IRLS procedure produces a solution to \eqref{eq:loglinearG} by utilizing a quadratic approximation to the actual quasi-likelihood at the maximizing value; by standard theory the solution is an $o(1)$ approximation to the actual maximizing value of the quasi-likelihood.  This strategy appeals to the common approach of defining a quasi-likelihood from estimating equations by considering the dual quadratic minimization problem (see, for example, \citealt{Green1984,McCullagh1991}).

\subsection{The quasi-likelihood information criterion}

We now address selection of the blip model. Based on the preceding derived quasi-likelihoods, we propose an information criterion whose general form builds on standard likelihood theory, where the Kullback-Leibler divergence between a proposed model and the true, data-generating model is minimized.

Under standard regularity conditions on the quasi-likelihood function $Q(.)$, inference proceeds in the usual way for misspecified models.  Let $f(y)$ denote the true, data-generating distribution, and let $\gamma(y;\bm{\psi}_{(m)})$ denote a proposed blip model which, combined with treatment and treatment-free models, fully specifies the G-estimating quasi-likelihood, $Q(y;\bm{\psi}_{(m)})$, and the corresponding density $f_m(y) \equiv f(y;\bm{\psi}_{(m)})$.  The proposed blip model is taken from a class of candidate models, $\mathcal{M}(m) = \{\gamma(y;\bm{\psi}_{(m)})|\bm{\psi}_{(m)} \in \Psi(m)\}$ with fitted models $\gamma(y;\hat{\bm{\psi}}_{(m)})$. The divergence between $f(y)$ and $f_m(y)$ estimated using the observed data and $\hat{\bm{\psi}}_{(m)}$ is given (up to an additive constant) by $\delta(\hat{\bm{\psi}}_{(m)}) = E[-2Q(Y;\bm{\psi})]|_{\bm{\psi}_{(m)} = \hat{\bm{\psi}}_{(m)}}$, computed with $\hat{\bm{\psi}}_{(m)}$ fixed, and the expected divergence is given by $\Delta(m) = E[\delta(\hat{\bm{\psi}}_{(m)})]$; in the latter expression, the expectation is over the distribution of the estimator $\hat{\bm{\psi}}_{(m)}$.  All expectations are taken with respect to the true distribution $f(y)$ by considering independent copies of the data. 

Let $\bm{\psi}_{(m,*)} = \arg\min_{\bm{\psi}_{(m)} \in \Psi(m)}Q(y;\bm{\psi}_{(m)})$. If the true blip function is parametric and contained in $\mathcal{M}(m)$, then $\bm{\psi}_{(m,*)}$ is the ``true" parameter; if the set of candidate models does not contain the true blip, then $\bm{\psi}_{(m,*)}$ is the value such that $Q(y;\bm{\psi}_{(m,*)})$ provides the best approximation to $f(y)$ in the sense of minimizing the expected Kullback-Leibler divergence.  Under standard regularity conditions on $Q(.)$, we have that $\hat{\bm{\psi}}_{(m)}$ is consistent for $\bm{\psi}_{(m,*)}$, and
\[
\sqrt{n} (\hat{\bm{\psi}}_{(m)} - \bm{\psi}_{(m,*)}) \stackrel{d}{\longrightarrow} \text{Normal}(0,\mathcal{V}(\bm{\psi}_{(m,*)}))
\]
where $\mathcal{V}$ is a positive definite matrix given by $\mathcal{V}(\bm{\psi}) = \mathcal{I}(\bm{\psi})^{-1} \mathcal{J}(\bm{\psi}) \mathcal{I}(\bm{\psi})^{-1}$ where, for $\bm{\psi}^\prime \in \mathcal{N}$, an open neighborhood of $\bm{\psi}_{(m,*)}$
\[
\mathcal{I}(\bm{\psi}^\prime) = \left. E \left[-\frac{\partial^2 Q_1(\bm{\psi})}{\partial \bm{\psi} \partial \bm{\psi}^\top } \right] \right|_{\bm{\psi} = \bm{\psi}^\prime} \qquad \mathcal{J}(\bm{\psi}^\prime) = \left. E \left[ \left\{\frac{\partial Q_1(\bm{\psi})}{\partial \bm{\psi} }\right\}  \left\{ \frac{\partial Q_1(\bm{\psi})}{\partial \bm{\psi} } \right\}^\top \right] \right|_{\bm{\psi} = \bm{\psi}^\prime}.
\]
and
\[
\frac{\partial Q_1(\bm{\psi})}{\partial \bm{\psi}} = D_1 A_1 \bm{h}_{\psi 1} ({Y}_1^* - \mu_1^*(\bm{\psi}))
\]
is the G-estimating function inspired by \eqref{eq:poiseq2a} for the first data point.
\medskip

\noindent \textcolor{black}{\textbf{Theorem}: Suppose that $Q(\bm{\psi})$ is twice continuously differentiable with bounded expectation of its second derivative in an open neighborhood $\mathcal{N}$ of $\bm{\psi}_{(m,*)}$. Then, under the stable unit treatment value and no unmeasured confounding assumptions (detailed in Subsection 2.1), the expected divergence $\Delta(m)$ can be approximated as
\begin{align*}
\Delta(m) & =  E[-2Q(\bm{\psi}_{(m,*)})] + 2 \text{tr} \left\{\mathcal{J}(\bm{\psi}_{(m,*)}) \mathcal{I}(\bm{\psi}_{(m,*)})^{-1}  \right\} + o(1)
\end{align*}
which is consistently estimated by
\[
\text{QIC}_{G}(m) = \widehat \Delta(m) = -2Q(\hat{\bm{\psi}}_{(m)}) + 2 \text{tr} \{\bm{J}(\hat{\bm{\psi}}_{(m)}) \bm{I}(\hat{\bm{\psi}}_{(m)})^{-1}\}
\]
where $\bm{I}(.)$ and $\bm{J}(.)$ are the observed (empirical) versions of $\mathcal{I}$ and $\mathcal{J}$.   Thus, the model selection procedure that chooses a model by minimizing $\text{QIC}_{G}(m)$ across $\mathcal{M}(m)$ identifies the model that minimizes $\Delta(m)$ with probability 1 as $n \longrightarrow \infty$.}

\noindent \textbf{Proof}:  see Supplementary Material.

\noindent This result gives rise to our quasi-likelihood information criterion, which in terms of the estimator of the asymptotic variance $\mathcal{V}$, $\widehat{\bm{V}}(\hat{\bm{\psi}}) = n \bm{I}(\hat{\bm{\psi}})^{-1} \bm{J}(\hat{\bm{\psi}}) \bm{I}(\hat{\bm{\psi}})^{-1}$, may be written
\begin{eqnarray}\label{eq:QICG}
\text{QIC}_\text{G} = -2Q(\hat{\bm{\psi}}) + 2\text{tr}\{\bm{I}(\hat{\bm{\psi}}) \widehat{\bm{V}}(\hat{\bm{\psi}}))\}.
\end{eqnarray}
In their derivation of a related criterion, \cite{Taguri2014QIC} use a direct sandwich estimator for $\mathcal{V}$.  However, while this allows a slight simplification of expression (\ref{eq:QICG}), we note that this approach fails to accommodate all sources of uncertainty. The estimation of the parameters ${\bm{\alpha}}$ of the treatment model at each interval should be acknowledged, and as we move through stages recursively estimation of all previous parameters should be similarly accommodated; this is achieved through the application of Taylor expansions to the estimating function ${U}(\bm{\psi})$ \citep{ROBINS2004PROC,MOODIE2009GESTVAR}, although in our experience such corrections make little difference to the resulting variance estimates.

Our derivation of the QIC also differs from that of \cite{Taguri2014QIC} in two substantial ways.  First, the estimation of $\hat{\bm{\beta}}$ is corrected for automatically by its substitution in the estimation of $\hat{\bm{\psi}}$, that is, using implicit forms $\bm{\beta}(\bm{\psi})$ in the linear model or the IRLS recursion for the log-linear model.  That is, we do not estimate the treatment-free model parameters $\bm{\beta}$ in a separate calculation using only the untreated individuals. Secondly, our derivation of the quasi-likelihood matches that of \cite{Taguri2014QIC} in the linear case, however their approach cannot be extended to the log-linear case.

The form of (\ref{eq:QICG}) is typical in information criterion-style approaches \citep{Takeuchi76}, and writing $K = \text{tr}\{\bm{I}(\hat{\bm{\psi}}) \widehat{\bm{V}}(\hat{\bm{\psi}}))\}$ we may present it as $\text{QIC}_\text{G} = -2Q(\hat{\bm{\psi}}) + 2K$ to more clearly evoke this similarity. This criterion may be applied at each stage of the G-estimation process with the blip model returning the lowest criterion value being recommended, as in a more typical analysis. Note, however, that it is necessary to assume that at all but the first stage of treatment an at-worst overspecified blip model is contained within the set of candidate models as otherwise poor parameter estimation can have a cumulative effect. Similarly, we must assume that at least one of the treatment or treatment-free models is correctly specified, so that the resulting blip parameter estimators are consistent. These assumptions are necessary for any recursive procedure.

The above theory extends to the case of continuous treatments. The primary complication is that the blip function is extended to include a quadratic treatment term, so that the optimal treatment at any given stage may lie inside the range of possible values it may take \citep{RICH2014CT}. After this modification, we can proceed to define an equivalent quasi-likelihood (and quasi-likelihood information criterion) at each stage of treatment. Full details are included in the Supplementary Material.

\section{Analysis} \label{sec:Analysis}

In this section, we first use simulations to demonstrate the IRLS approach to G-estimation for a count outcome and to demonstrate the performance of the QIC$_G$ in a continuous outcome scenario. We then proceed to apply both IRLS and the quasi-likelihood information criterion to an empirical analysis, performing analyses which treat the (discrete) outcome as either continuous or as a count.

\subsection{Simulation study: IRLS for a log-linear SNMM}

First, we present an illustration of the IRLS algorithm for G-estimation in the case of a log-linear outcome model. Simulating a two-stage example, we generate data as follows:

\begin{itemize}
\item stage 1 patient information: $X_{1} \sim N(0,1)$;
\item stage 1 treatment: $a_1 \in \{0,1\}$, ${P}(A_1 = 1 | \bm{h}_1) = \text{expit}(x_{1})$;
\item stage 2 patient information: $X_{2} \sim N(a_1,1)$;
\item stage 2 treatment: $a_2 \in \{0,1\}$, ${P}(A_2 = 1 | \bm{h}_2) = \text{expit}(x_{2})$;
\item stage $j$ blip: $\gamma_j(a_j,x_j) = a_j(\psi_{j0} + \psi_{j1} x_{j1})$ such that $a_j^{opt} = {1}_{\{\psi_{j0} + \psi_{j1} x_{j1} > 0\}}$;
\item outcome: $P(Y = k) = \lambda^k e^{-k}/k!$,
\subitem with $\lambda = \exp\left[\beta_0 + \log(|x_1|) - \sum_{j=1}^2[\gamma_j(a_j^{opt},x_j) - \gamma_j(a_j,x_j)]\right]$.
\end{itemize}
For all simulations we set $(\psi_{j0},\psi_{j1}) = (0.5,-0.5)$, $j = 1, 2$. As described in section \ref{sec:Multiplicative}, one concern in extending G-estimation to discrete outcomes is that of zero values in the response, and the effect they can have on the stage-specific pseudo-outcomes. Replacing the 0 with 0.001 when computing the pseudo-outcome, we investigate the performance of the algorithm in three sets of simulations with varying values for $\beta_0$, chosen to yield outcomes with approximately 5\%, 10\% and 20\% zeros. Our analyses correctly specified the treatment model, but mis-specified the treatment-free model, supposing it was linear in $x_1$ in contrast to the true $\log(|x_1|)$ term.  Initial simulation runs had unacceptably high rates of failure to converge. To address this we adjusted the IRLS algorithm detailed above slightly, introducing step-halving whereby the initial parameter estimates at each iteration were the mean of the previous two stages, and ignoring the termination condition dependent on $|\bm{\mu}^{(t)} - \bm{\mu}^{(t-1)}|$, instead terminating when only $|\bm{\eta}^{(t)} - \bm{\eta}^{(t-1)}|$ dropped below a given tolerance. This reduced convergence failure rates to 3\% or lower.

We generated 1000 simulated datasets per setup, setting the tolerance $\epsilon_\eta$ to 0.001 and limiting the number of iterations at each stage to 1000. Exploratory analyses of smaller simulation runs with lower tolerances and larger iteration limits did not yield substantially different results for parameter estimates or failure rates. Results are summarized (Table \ref{tab:PoisSims}), where stage 1 estimates for the covariate-by-treatment interaction are slightly (though not statistically significantly) biased in small samples. Bias does not appear to be related to the probability of zero-outcomes, although standard errors appear to increase with it.

\begin{table}[ht]
\centering
\caption{Mean blip parameter estimates (standard errors) from 1000 simulation runs for log-linear outcome model via iteratively-reweighted least squares. True blip parameters $(\psi_{j0},\psi_{j1})$ of $(0.5,-0.5)$.  }\label{tab:PoisSims}
\begin{tabular}{rrrrrrr}
\hline
 $n$ & $P(Y = 0)$  & $\widehat{\psi}_{10}$ (SE) & $\widehat{\psi}_{11}$  (SE) & $\widehat{\psi}_{20}$  (SE) & $\widehat{\psi}_{21}$  (SE) \\\hline
50    & 5\%  & 0.485 (0.337) & -0.349 (0.260) & 0.487 (0.320) & -0.437 (0.372) \\
       & 10\% &  0.511 (0.370) & -0.345 (0.295) & 0.500 (0.348) & -0.449 (0.407)\\
       & 20\% &  0.523 (0.418) & -0.339 (0.343) & 0.496 (0.388) & -0.445 (0.461)\\  
100    & 5\%  & 0.503 (0.209) & -0.427 (0.155) & 0.497 (0.217) & -0.475 (0.235) \\
       & 10\% &  0.501 (0.226) & -0.428 (0.165) & 0.496 (0.232) & -0.470 (0.247)\\
       & 20\% &  0.520 (0.248) & -0.419 (0.191) & 0.504 (0.266) & -0.483 (0.287)\\  
200    & 5\%  & 0.501 (0.144) & -0.471 (0.101) & 0.504 (0.153) & -0.486 (0.166) \\
       & 10\% & 0.501 (0.154) & -0.471 (0.109) & 0.502 (0.163) & -0.486 (0.171) \\
       & 20\% & 0.510 (0.174) & -0.470 (0.124) & 0.501 (0.186) & -0.484 (0.198)  \\ 
500    & 5\%  & 0.498 (0.089) & -0.485 (0.062) & 0.502 (0.095) & -0.496 (0.102)\\
       & 10\% & 0.497 (0.095) & -0.485 (0.066) & 0.505 (0.100) & -0.495 (0.110) \\
       & 20\% & 0.502 (0.107) & -0.485 (0.077) & 0.500 (0.114) & -0.495 (0.120) \\\hline
\end{tabular}
\end{table}

\subsection{Simulation study: QIC$_G$}

Next, we demonstrate the use of QIC$_G$ in the DTR framework with a variety of simulated two-stage examples from the continuous outcome setting (we present results for the discrete-outcome setting in the Supplementary Material). We generate data as follows:

\begin{itemize}
\item stage 1 patient information: $X_{1k} \sim N(0,1)$ for $k = 1, 2, 3$;
\item stage 1 treatment: $a_1 \in \{0,1\}$, ${P}(A_1 = 1 | \bm{h}_1) = \text{expit}(x_{11}+x_{12}+x_{13})$;
\item stage 2 patient information: $X_{2k} \sim N(a_1,1)$ for $k = 1, 2, 3$;
\item stage 2 treatment: $a_2 \in \{0,1\}$, ${P}(A_2 = 1 | \bm{h}_2) = \text{expit}(x_{21}+x_{22}+x_{23})$;
\item stage $j$ blip: $\gamma_j(a_j,\bm{h}_j) = a_j(1 + \psi_{j1} x_{j1} + \psi_{j2} x_{j2} + \psi_{j3} x_{j3})$
\subitem such that $a_j^{opt} = 1_{\{1 + \psi_{j1} x_{j1} + \psi_{j2} x_{j2} + \psi_{j3} x_{j3} > 0\}}$;
\item outcome: $Y = -\sum_{j=1}^2[\gamma_j(a_j^{opt},\bm{h}_j) - \gamma_j(a_j,\bm{h}_j)] + \epsilon$, with $\epsilon \sim \text{log-normal}(0,1) - e^{0.5}$;
\end{itemize}

\noindent where $\text{expit}(x) = \left[1+\text{exp}(-x)\right]^{-1}$ is the expit or inverse-logit function. We have used skewed errors in our generation of the outcome (centralized to have mean zero) to better illustrate the potential benefits of the QIC$_G$ approach. Results using normal errors are included in the Supplementary Material for reference. In our first analyses, we consider datasets of size $n = 50, 100$ and $200$, and set the blip parameters to $(\psi_{j1},\psi_{j2},\psi_{j3}) = (1,0,0)$, $(1,1,0)$ or $(1,1,1)$ giving a range of models including one, two, or all three variables at each stage.

We conducted a G-estimation analysis of 1000 simulated datasets considering eight different blip models corresponding to each of the possible combinations of the predictors at each stage (that is, using none, one, two or all three). The treatment models were always correctly specified (and modeled using logistic regression), while the treatment-free models $E[{G}_j (\bm{\underline{\psi}}_j)|\bm{h}_{\beta j};\bm{\beta}_j]$ were linear with covariates $(1,x_{11},x_{12},x_{13})$ at stage 1 and $(1,x_{11},x_{12},x_{13},a_1x_{11},a_1x_{12},a_1x_{13},x_{21},x_{22},x_{23})$ at stage 2 (i.e. using all available covariates).

Using QIC$_G$, we performed forward and backward selection within the set of candidate models as in a standard AIC-type stepwise analysis. For comparison we also conducted forward and backward selection based on Wald test p-values at a 0.05 significance level. For these initial simulations, stage 1 results are based on analysis carried out following fitting of the correct model at stage 2. Results (Table \ref{tab:table1}) indicate QIC$_G$ outperforms the Wald-type approaches except for the smallest true models (although even then it largely remains competitive). Furthermore, while QIC$_G$ shows a slight tendency to overfit, the Wald-type approaches show considerably more bias towards underfitting. In addition, we note a slightly greater consistency between the forward and backward QIC$_G$ results than between the Wald test results, suggesting QIC$_G$ may be more robust to choice of selection direction.

\begin{table}[ht]
\centering
\caption{Model selection for a variety of sample sizes ($n$). Numbers indicate proportion of 1000 simulation runs where the correct model was selected by the corresponding method. (F) and (B) denote forwards and backwards selection, respectively. \textbf{Bold} indicates the most successful approach for each setup.} \label{tab:table1}
\begin{tabular}{llcccc}
  \hline
$n$ & Model & QIC$_G$ (F) & QIC$_G$ (B) & Wald (F) & Wald (B) \\
  \hline
50 & $x_{11}$ & 0.285 & 0.276 & \textBF{0.303} & 0.297 \\
   & $x_{11},x_{12}$ & 0.232 & \textBF{0.238} & 0.137 & 0.160\\
   & $x_{11},x_{12},x_{13}$ & 0.218 & \textBF{0.259} & 0.060 & 0.108 \\
   & $x_{21}$ & 0.225 & 0.216 & \textBF{0.263} & 0.248 \\
   & $x_{21},x_{22}$ & 0.203 & \textBF{0.213} & 0.119 & 0.145 \\
   & $x_{21},x_{22},x_{23}$ & 0.181 & \textBF{0.214} & 0.068 & 0.09 \\
  100 & $x_{11}$ & 0.437 & 0.423 & \textBF{0.479} & 0.471 \\
   & $x_{11},x_{12}$ & 0.373 & \textBF{0.380} & 0.284 & 0.318 \\
   & $x_{11},x_{12},x_{13}$ & 0.372 & \textBF{0.419} & 0.181 & 0.254 \\
   & $x_{21}$ & 0.372 & 0.366 & \textBF{0.422} & 0.418 \\
   & $x_{21},x_{22}$ & 0.353 & \textBF{0.365} & 0.263 & 0.285 \\
   & $x_{21},x_{22},x_{23}$ & 0.357 & \textBF{0.381} & 0.150 & 0.205 \\
  200 & $x_{11}$ & 0.533 & 0.530 & \textBF{0.606} & 0.601 \\
   & $x_{11},x_{12}$ & 0.569 & \textBF{0.573} & 0.484 & 0.519 \\
   & $x_{11},x_{12},x_{13}$ & 0.587 & \textBF{0.628} & 0.379 & 0.451 \\
   & $x_{21}$ & 0.479 & 0.477 & \textBF{0.581} & 0.577 \\
   & $x_{21},x_{22}$ & 0.505 & \textBF{0.507} & 0.439 & 0.453 \\
   & $x_{21},x_{22},x_{23}$ & 0.538 & \textBF{0.557} & 0.334 & 0.371 \\
   \hline
\end{tabular}
\end{table}

In these multi-stage simulations, all methods perform better in selection of the first stage blip model rather than the second due to the first stage being a simpler model fitting problem: at stage 2 both first and second stage covariates affect analysis. We note however that the recursive nature of the G-estimation approach means that the analysis of the first stage happens subsequent to analysis of the second stage, using estimates obtained from that stage which we might expect to affect stage 1 model assessment. It seems that any impact this aspect of the estimation process might have is small compared to the inherent simplicity that earlier stages involve fewer covariates.

Moreover, in the results for stage 1 model assessment we have ignored the problem of stage 2 model selection, and instead fixed the stage 2 model as the correct one. While in reality this is an implicit assumption we must make, it seems prudent to investigate what impact this may have on model selection. We conduct analyses identical to those above (limited to $n = 100$) but with two other approaches to model selection. In addition to the `best-case' scenario of selecting the correct stage 2 model, we also investigate the consequences of choosing the model recommended by the model selection procedures themselves, and a `worst-case' scenario where intercept-only models are used (Table \ref{tab:table2}). These non-optimal approaches do result in a drop in performance, but this is not particularly dramatic (and not statistically significant) even in the worst-case scenario. We note, however, that in more complex setups it is likely mis-specification of the stage 2 models could have more dramatic consequences for stage 1 model selection. 

\begin{table}[ht]
\centering
\caption{Stage 1 model selection ($n = 100$) when the stage 2 model is correctly specified (`Correct'), selected by the corresponding approach (`Recommended'), or an intercept-only model (`Intercept'). Numbers indicate proportion of 1000 simulation runs where the correct model was selected by the corresponding method. (F) and (B) denote forwards and backwards selection, respectively. \textbf{Bold} indicates the most successful approach for each setup.} \label{tab:table2}
\begin{tabular}{llcccc}
  \hline
Selection & Model & QIC$_G$ (F) & QIC$_G$ (B) & Wald (F) & Wald (B) \\
  \hline
Correct & $x_{11}$ & 0.437 & 0.423 & \textBF{0.479} & 0.471 \\
   & $x_{11},x_{12}$ & 0.373 & \textBF{0.380} & 0.284 & 0.318 \\
   & $x_{11},x_{12},x_{13}$ & 0.372 & \textBF{0.419} & 0.181 & 0.254 \\
  Recommended & $x_{11}$ & 0.420 & 0.409 & \textBF{0.485} & 0.471 \\
   & $x_{11},x_{12}$ & 0.369 & \textBF{0.375} & 0.283 & 0.316 \\
   & $x_{11},x_{12},x_{13}$ & 0.372 & \textBF{0.417} & 0.175 & 0.250 \\
  Intercept & $x_{11}$ & 0.431 & 0.417 & \textBF{0.479} & 0.469 \\
   & $x_{11},x_{12}$ & 0.373 & \textBF{0.380} & 0.283 & 0.311 \\
   & $x_{11},x_{12},x_{13}$ & 0.373 & \textBF{0.411} & 0.167 & 0.245 \\
   \hline
\end{tabular}
\end{table}

We also investigated the impact of other aspects of our data generation setup. Weaker effect sizes (simulated by setting the blip parameters to 0.1 and 0.5 instead of 1 as above) predictably resulted in lower success rates across all the methods under consideration. However, the quasi-likelihood approach was much more resilient to these effects. Introducing a correlation structure among the non-treatment covariates also resulted in worse model selection, but again the quasi-likelihood approach appeared to be slightly more robust to these changes. Results of these additional scenarios are included in the Supplementary Material.

These results were aggregated across the various model setups at each stage to afford greater simplicity in the presentation of our results. For example, the results corresponding to the stage 1 blip model which only included $x_{11}$ were taken from simulations across the three different stage 2 blip models. Non-aggregated results (see Supplementary Material) show little evidence of stage 2 model complexity affecting stage 1 selection or vice-versa.

\subsection{Investigating the trace term}  We have framed our $\text{QIC}_{G}$ in terms of the quasi-likelihood and a `penalty term' defined in terms of the trace $K = \text{tr}\{\bm{I}(\hat{\bm{\psi}})\widehat{\bm{V}}(\hat{\bm{\psi}})\}$. In the likelihood-based setting it is known that this trace term may, if the model under consideration is a good approximation of the truth, be approximated by the dimension of that model \citep{BurnhamAnderson,ShibataAIC}. Here, the quasi-likelihood is grounded in estimating equation theory and so, if the error terms in the outcome generating model were normally distributed, we might expect a similar result. This is complicated by our estimation of two other models besides the blip, as well as the recursive multi-stage nature of the G-estimation framework.

We summarize estimates of the term $K$ from our first set of simulations above with $n = 100$ in a figure in the supplement, with the modification that we generate the error term $\epsilon$ in the outcome from a standard normal distribution rather than log-normal. In general, the estimates appear to be similar to the dimension of the corresponding model, particularly at the second stage, when the correct model was used. If two models of the same dimension are compared then on average an incorrect model will result in a slightly larger trace term than that from a correct model. Stage 1 estimates of the trace in general seem to be slightly lower. These results assume correct specification of the treatment model, ensuring consistency of our estimators as per the double-robustness property of G-estimation. We have found (results omitted) that when both treatment and treatment-free models are mis-specified the resulting trace terms can be much larger. In addition, if the distribution of the error term $\epsilon$ is skewed, as in the previously reported results, then the trace term is again much larger. This suggests the possibility of comparing the trace term with candidate model dimension to investigate the validity of the treatment and treatment-free models, following use of residual plots (or similar techniques) to assess the normality of the residuals.

We note finally that the use of bootstrap procedures to estimate the trace (or penalty) term $\text{tr}\{\bm{I}(\hat{\bm{\psi}})\widehat{\bm{V}}(\hat{\bm{\psi}})\}$ have been recommended in preference to the use of termwise plug-in versions of matrices $\bm{I}(.)$ and $\bm{J}(.)$ (see \citet{BurnhamAnderson}).  We have not investigated this possibility in our analysis as the plug-in procedure appears to work well in our examples, and due to the additional computational burden.   Bootstrap procedures are valid for inference in the regular G-estimation setting (see for example \cite{SHORTREED2012SCHIZ}), and are straightforward, although computationally expensive, to implement.

\subsection{Sequenced Treatment Alternatives to Relieve Depression study}

We now illustrate application of our proposed IRLS and QIC$_G$ approaches to real data from the Sequenced Treatment Alternatives to Relieve Depression (STAR*D) study. STAR*D was a multi-stage randomized control trial designed to compare different treatment regimes for patients with major depressive disorder \citep{FAVA2003STARD,RUSH2004STARD}. The study was split into 4 levels (one of which was itself split into two sub-levels), with patients receiving a different treatment or combination of treatments within each level.

At study entry (level 1) patients were prescribed citalopram and followed up at regularly scheduled clinic visits. Those whose depression did not enter remission -- defined as a Quick Inventory of Depressive Symptomatology (QIDS) score less than or equal to 5 -- could proceed to a second level of treatment where seven treatment options were available. The second level of treatment was characterized by `switching' from citalopram to one of four new treatments, or `augmenting' the current treatment by receiving citalopram alongside one of three other treatment options. Patients who received cognitive therapy at level 2 (either alone or combined with citalopram) were eligible to enter the sublevel 2A where they received one of the treatments available at level 2. All patients without remission could then proceed to level 3 (and, if their depression persisted, a further level 4) where again their previous treatment was either switched to or augmented with a number of options. Full details of the study design and treatment options, are described elsewhere \citep{RUSH2004STARD}.

An important aspect of the study is that patients were asked for their treatment \emph{preference} and would then be randomized to one of the treatment options consistent with their preference. This is typically characterized as patients choosing to `switch' from their current treatment to, or `augment' it with, a different one, although the reality was slightly more complex. In particular, when moving from level 1 to level 2, patients were asked about their preference to switch or augment their current treatment with cognitive therapy separately to their preference to switch or augmenting with a pharmacological (i.e., drug-based) treatment.


We conduct an analysis following \citet{CHAK2013STARD} who investigated the dichotomy between treatments that were, or included, a selective serotonin reuptake inhibitor (SSRI) and those that did not. We restrict attention to two stages of the study and consider level 2 (including level 2A) and level 3 as our first and second stages of treatment, respectively. Treatment at each stage was coded as 1 if an SSRI was received, either alone or in combination, with level 2A treatments (both of which were non-SSRI) combined with level 2 treatments for this purpose. Treatment was coded as 0 if no SSRI was received throughout a stage. Of 1,027 total patients, only 273 entered level 3. Our outcome is defined as negative QIDS score at end of treatment (i.e., at the end of stage 2 if a patient entered level 3, and at the end of stage 1 otherwise). By taking the negative, larger values are preferred, and we therefore seek a DTR that maximizes this outcome. We pursue an analysis analogous to those undertaken by previous authors, viewing QIDS score as a continuous outcome, and then use our new IRLS-based approach and the associated QIC$_G$ to apply a log-linear model.

We consider the following tailoring variables: QIDS score measured at the start of the corresponding level (denoted $q_j$ for stage $j$), the change in QIDS score divided by time across the previous level (\emph{QIDS slope}, denoted $s_{j}$), and patient preference prior to receiving treatment ($p_j$). Patient preference is binary and coded as 1 if the patient rejected all treatments consistent with switching to a different pharmacological treatment, and 0 otherwise. The assumed treatment model at each stage was fit by logistic regression of observed treatment on preference only, while the treatment-free models are specified as:

\begin{itemize}
\item stage 1: $E[{G}_1 (\bm{\underline{\psi}}_1)|\bm{h}_{\beta 1};\bm{\beta}_1] = \beta_{10} + q_1\beta_{11} + s_1\beta_{12} + p_1 \beta_{13}$; and
\item stage 2: $E[{G}_2 (\bm{\underline{\psi}}_2)|\bm{h}_{\beta 2};\bm{\beta}_2] = \beta_{20} + q_{2}\beta_{21} + s_{2}\beta_{22} + p_{2} \beta_{23} + a_1 \beta_{24}$.
\end{itemize}

We consider the `full' blip models

\begin{itemize}
\item stage 1: $\bm{\gamma}_1(\bm{h}_{\psi 1},{a}_1;\bm{\psi}_1) = a_1(\psi_{10} + q_{1}\psi_{11} + s_{1} \psi_{12} + p_{1} \psi_{13})$; and
\item stage 2: $\bm{\gamma}_2(\bm{h}_{\psi 2},{a}_2;\bm{\psi}_2) = a_2(\psi_{20} + q_{2}\psi_{21} + s_{2} \psi_{22})$,
\end{itemize}

\noindent and investigate all sub-models (every covariate combination from the full models, including intercept-only models). There are eight candidate models at stage 1, and four at stage 2.


Rather than proceed directly to a stepwise procedure, we instead fit all possible models as dimensionality was low. G-estimation was therefore first performed four times: once for each of the candidate stage 2 blip models, with an intercept-only blip model specified at stage 1 (this choice not affecting stage 2 analysis). The lowest value of $\text{QIC}_{G}$ was found when the intercept-only model was used, suggesting it is the best choice of stage 2 blip model (based on either a forward or backward selection procedure). This result was reinforced by application of both Wald-type approaches of the previous section.

We then repeated the analysis for each of the eight candidate stage 1 blip models using the recommended stage 2 blip model, and found that the model containing stage 1 preference only returned the lowest QIC$_G$ (and would be recommended by either a forward or backward procedure). The Wald-type approaches also both recommended the preference-only model.

Our log-linear analyses were based on a similar setup with treatment-free and blip models exponentiated, adding 27 to all outcome measures to ensure they were positive. Applying G-estimation via IRLS and computing QIC$_G$ for every blip model we found the same models recommended at each stage. We observe that in all analyses the same model was recommended at stage 1 regardless of which stage 2 model was chosen.

Overall, these results are broadly consistent with the analysis of Chakraborty et al. who, despite using the somewhat different approach of Q-learning, found that no stage 2 blip covariates were statistically significant, while stage 1 preference alone was significant in the stage 1 blip model. It is encouraging that QIC$_G$ for both modeling setups indicated the preference term should be included.  We also find that at both stages both estimation processes predict the same optimal treatment for every patient.

Finally, we can compare observed outcomes among patients based on how consistent their observed treatments were with the optimal ones as recommended by our models. Among patients who entered stage 2, mean improvement in QIDS score among those who received optimal treatment at both stages, one stage, or no stages was, respectively, 4.67 (sd = 3.92, $n$ = 15), 3.80 (sd = 6.08, $n$ = 75), and 2.56 (sd = 5.19, $n$ = 183). Among patients who did not enter stage 2, mean improvement was 6.13 (sd = 4.72, $n$ = 145) for those who received optimal treatment, and 5.49 (sd = 4.84, $n$ = 609) for those who did not.

\section{Discussion}

Personalized medicine and the development of dynamic treatment regimes is an important frontier in biostatistical research. This has been reflected in a rapidly expanding literature focusing on DTR estimation techniques, but more practical concerns have received comparatively little attention. In this paper we have presented two extensions to the G-estimation framework. First, we have demonstrated how G-estimation may be applied for log-linear models via iteratively-reweighted least squares; this provides a relatively straightforward route to G-estimation use in a greater variety of contexts. Further, we have presented an approach to model selection for SNMMs within the DTR framework. By demonstrating how G-estimation in its typical application (for continuous treatments) may be reduced to a relatively simple form, we have derived a quasi-likelihood for each stage of a multi-stage, recursive analysis process. We have then extended the work of \citet{Pan2001QIC} and \citet{Taguri2014QIC} to derive a general quasi-likelihood information criterion for DTR estimation using G-estimation. Furthermore, while we have focused on the binary treatment setting, the theory extends to the case of a continuous treatment, dramatically increasing its applicability.

Our simulations involving log-linear SNMMs indicate how G-estimation may be implemented for discrete outcomes with relative ease, especially if extant IRLS routines in standard software packages can be used for this purpose. We note that there may be further room to improve on the approximation used to handle the zero-outcomes, which at present are set to a very small number only when the non-optimal treatment was received. This concern extends to the binary outcome setting, where estimation is even more problematic: the `blip' cannot be separated from the treatment-free component of the mean using a logit transformation, and the use of a log-linear model only provides reasonable (unbiased) estimators in a small range of settings. The question of how to construct pseudo-outcomes that better address this issue is an avenue for further research.   Note, finally, that if the (counterfactual) outcomes are binary, then modeling the expected counterfactual on the usual logistic scale is problematic for G-estimation, as it is not possible to separate the blip component from the treatment-free expected counterfactual component, as would be the case on the linear or log-linear scale.  This issue warrants further attention.

Through simulation studies we have shown that our quasi-likelihood information criterion performs as well as or better than simpler Wald-type approaches for continuous outcomes, particularly when sample or effect sizes are small, or there is correlation between candidate covariates. In addition, we found greater agreement between the forward and backward stepwise approaches when using QIC$_G$ than the Wald-type approach, a potentially attractive feature in practice. We note, however, that QIC$_G$ does seem to overfit, and as such slight modifications may lead to more balanced results. We have experimented with ad hoc corrections inspired by the Bayesian Information Criterion and the corrected AIC, which have yielded promising results. Moreover while one may often argue that for the purposes of an explanatory analysis, compared with underfitting, overfitting is the less serious error, this is perhaps more justified in a multi-stage setting where mis-specification by underfitting can have a more severe knock-on effect in our analysis.  When the outcome was generated using normal errors, rather than the skewed log-normal errors discussed here, the Wald-type approaches become more competitive, but still badly underfit (full results are included in the Supplementary Material). As in any simulation-based analysis we appreciate that the results presented here cannot possibly be comprehensive, and so would encourage further experimentation (and analysis) to assess the properties of these various criteria. Of particular interest is a more extensive investigation of the QIC$_G$ for discrete outcomes, with our preliminary analyses providing encouraging results.

The trace term, $K$, warrants additional research. In our simulations we observed that when errors were normally distributed, and the blip model correctly specified, $K$ was approximately equal to the dimension of the model (as is the case in the likelihood-based setting). This was not generally the case, however, especially when both the treatment and treatment-free models were badly mis-specified or errors were non-normal. From a practical perspective, the quasi-likelihood criteria presented may tend to underfit when models are mis-specified (as the corresponding penalty term is large), but if the researcher has doubts about the legitimacy of parameter estimators (due to mis-specification of both the treatment and treatment-free models), then model selection is a secondary concern. In our analysis of the STAR*D data we found that for the various candidate models, the trace term was somewhat larger than the dimension of the associated blip function while residual plots from the proposed models were consistent with normal errors. This could indicate an inadequacy in the treatment or treatment-free models which might merit further investigation. 

Our QIC formulation has focused exclusively on selection of the blip, or contrast, component of the outcome mean model. However, the approach can easily be adapted to select the entire mean model (the contrast and treatment-free models simultaneously), allowing the use of our information criterion for G-estimation of static treatment sequences or G-estimation for mediation, as well as in binary outcome settings.

\section{Supplementary Material}

\begin{description}

\item[Appendix:] Appendix containing additional theory, and simulation results.
\item[R code:] R code for the simulation studies of section \ref{sec:Analysis} and the Appendix.

\end{description}

{\center{\Large{Appendix: Supplementary Material for Manuscript titled ``Generalized G-estimation and Model Selection''}}}

\section{Quasi-likelihood for continuous treatments}

In the binary treatment setting we showed that our estimating equations ${U}(\bm{\psi})$ may be reduced to
\begin{eqnarray*}
{U}(\bm{\psi}) &=& (\bm{D}\bm{h}_\psi)^\top \left[(\mathbf{I}_n - \hat{\bm{h}}_\beta) (\tilde{\bm{y}} - \bm{A} \bm{h}_\psi \bm{\psi})\right].
\end{eqnarray*}
\noindent We now extend to the case of a continuous treatment and include a quadratic term in our blip such that
\begin{eqnarray*}
\gamma(\bm{h}_\psi,{a};\bm{\psi}_1,\bm{\psi}_2) = {a} \bm{h}_{\psi_1} \bm{\psi}_1 + {a}^2 \bm{h}_{\psi_2} \bm{\psi}_2
\end{eqnarray*}
\noindent where we have compartmentalized our blip parameters $\bm{\psi} = (\bm{\psi}_1,\bm{\psi}_2)$ and history design matrix $\bm{h}_\psi = (\bm{h}_{\psi_1},\bm{h}_{\psi_2})$ depending on whether they are associated with the linear or quadratic term of ${a}$ in our blip. Writing $\bm{D}_1$ and $\bm{D}_2$ for the diagonal matrices with $(i,i)^{th}$ entry $a_i - {E}[A_i|H_i]$ and $a_i^2 - {E}[A_i^2|H_i]$, respectively, our estimating equations become \citep{RICH2014CT}
\begin{eqnarray*}
{U}(\psi) &=& \left( \begin{array}{c}
\bm{D}_1 \bm{h}_{\psi_1} \\
\bm{D}_2 \bm{h}_{\psi_2} \end{array} \right) ^\top \left[(\mathbf{I}_n - \hat{\bm{h}}_\beta) (\tilde{\bm{y}} - \bm{A} \bm{h}_{\psi_1} \bm{\psi}_1 - \bm{A}^2 \bm{h}_{\psi_2} \bm{\psi}_2)\right]
\end{eqnarray*},
\noindent which yields a quasi-likelihood of the form
\begin{eqnarray*}
Q(\bm{\psi}) &=& \bm{\psi}^\top \left( \begin{array}{c}
\bm{D}_1 \bm{h}_{\psi_1} \\
\bm{D}_2 \bm{h}_{\psi_2} \end{array} \right)^\top \left[(\mathbf{I}_n - \hat{\bm{h}}_\beta) \tilde{\bm {y}}\right] - \frac{1}{2}\bm{\psi}^\top \left( \begin{array}{c}
\bm{D}_1 \bm{h}_{\psi_1} \\
\bm{D}_2 \bm{h}_{\psi_2} \end{array} \right)^\top \left( \begin{array}{cc}
\bm{A} \bm{h}_{\psi_1} , \bm{A}^2 \bm{h}_{\psi_2} \end{array} \right)^\top \bm{\psi}\\
&=& \bm{\psi}^\top \bm{m}_c - \frac{1}{2}\bm{\psi}^\top \bm{M}_c \bm{\psi}
\end{eqnarray*}
\noindent where $\bm{m}_c$ and $\bm{M}_c$ may be thought of as the continuous treatment analogs to $\bm{m}$ and $\bm{M}$ derived in the main paper for the binary case.

\section{Proof of discrepancy theorem}

Recall the theorem from section 2.4 of the associated paper:

\noindent \textcolor{black}{\textbf{Theorem}: Suppose that $Q(\bm{\psi})$ is twice continuously differentiable with bounded expectation of its second derivative in a neighbourhood $\mathcal{N}$ of $\bm{\psi}_{(m,*)}$. Then, under the stable unit treatment value and no unmeasured confounding assumptions (detailed in section 2.1), the expected divergence $\Delta(m)$ can be approximated
\begin{align*}
\Delta(m) & =  E[-2Q(\bm{\psi}_{(m,*)})] + 2 \text{tr} \left\{\mathcal{J}(\bm{\psi}_{(m,*)}) \mathcal{I}(\bm{\psi}_{(m,*)})^{-1}  \right\} + o(1)
\end{align*}
which is consistently estimated by
\[
QIC_{G}(m) = \widehat \Delta(m) = -2Q(\hat{\bm{\psi}}_{(m)}) + 2 \text{tr} \{\bm{J}(\hat{\bm{\psi}}_{(m)}) \bm{I}(\hat{\bm{\psi}}_{(m)})^{-1}  \}
\]
where $\bm{I}(.)$ and $\bm{J}(.)$ are the observed (empirical) versions of $\mathcal{I}$ and $\mathcal{J}$.   Thus, the model selection procedure that chooses a model by minimizing $QIC_{G}(m)$ across $\mathcal{M}(m)$ identifies the model that minimizes $\Delta(m)$ with probability 1 as $n \longrightarrow \infty$.}

\bigskip

\noindent \textbf{Proof}: Following \citet{Takeuchi76} -- see also \citet{BurnhamAnderson} -- our quasi-likelihood information criterion (QIC) is based on an estimate of $\Delta(m)$, a function of the Kullback-Leibler discrepancy between the data generating model and model $m$.  Consider a decomposition of $\Delta(m) = E[\delta(\hat{\bm{\psi}}_{(m)})]$, where
\[
\delta(\hat{\bm{\psi}}_{(m)}) = E[-2Q(Y;\bm{\psi})]|_{\bm{\psi}_{(m)} = \hat{\bm{\psi}}_{(m)}},
\]
given by
\begin{align*}
\Delta(m) & = \left\{ E[\delta(\hat{\bm{\psi}}_{(m)})] - E[-2Q({\bm{\psi}}_{(m,*)}] \right\} \\[6pt]
& + \left\{  E[-2Q({\bm{\psi}}_{(m,*)})] - E[-2Q(\hat{\bm{\psi}}_{(m)})] \right\} + E[-2Q(\hat{\bm{\psi}}_{(m)})].
\end{align*}
We consider first an expansion of $Q(.)$ around the ``true" blip parameter for model $m$, ${\bm{\psi}}_{(m,*)}$: we have that
\begin{align*}
Q(\hat{\bm{\psi}}_{(m)}) & = Q({\bm{\psi}}_{(m,*)}) + \dot{Q}({\bm{\psi}}_{(m,*)}) (\hat{\bm{\psi}}_{(m)}-{\bm{\psi}}_{(m,*)}) \\[6pt]
& \qquad \qquad + \frac{1}{2} (\hat{\bm{\psi}}_{(m)}-{\bm{\psi}}_{(m,*)})^\top \ddot{Q}({\bm{\psi}}_{(m,*)}) (\hat{\bm{\psi}}_{(m)}-{\bm{\psi}}_{(m,*)}) + o_p(1)
\end{align*}
so taking expectations with respect to the data generating model with $\hat{\bm{\psi}}_{(m)}$ fixed, we have
\begin{align*}
E[-2Q(\hat{\bm{\psi}}_{(m)})] & = E[-2Q({\bm{\psi}}_{(m,*)})] - (\hat{\bm{\psi}}_{(m)}-{\bm{\psi}}_{(m,*)})^\top \mathcal{I}({\bm{\psi}}_{(m,*)}) (\hat{\bm{\psi}}_{(m)}-{\bm{\psi}}_{(m,*)}) + o(1)
\end{align*}
where, recall,
\[
\mathcal{I}(\bm{\psi}^\prime) = \left. E \left[-\frac{\partial^2 Q_1(\bm{\psi})}{\partial \bm{\psi} \partial \bm{\psi}^\top } \right] \right|_{\bm{\psi}= \bm{\psi}^\prime}
\]
and hence
\[
E[-2Q({\bm{\psi}}_{(m,*)})] - E[-2Q(\hat{\bm{\psi}}_{(m)}) ] = (\hat{\bm{\psi}}_{(m)}-{\bm{\psi}}_{(m,*)})^\top \mathcal{I}({\bm{\psi}}_{(m,*)}) (\hat{\bm{\psi}}_{(m)}-{\bm{\psi}}_{(m,*)}) + o(1).
\]
Using the conventional theory of misspecified models, we have that
\[
\sqrt{n}(\hat{\bm{\psi}}_{(m)}-{\bm{\psi}}_{(m,*)}) = \left\{\bm{I}({\bm{\psi}}_{(m,*)})\right\}^{-1} \times \frac{1}{\sqrt{n}} \sum_{i=1}^n \dot{Q}(Y_i;{\bm{\psi}}_{(m,*)}) + o_p(1)
\]
where, by construction of the quasi-likelihood function, we have
\[
\dot{Q}(y;{\bm{\psi}}) \equiv U(y;{\bm{\psi}}).
\]
Hence
\[
\sqrt{n}(\hat{\bm{\psi}}_{(m)} - {\bm{\psi}}_{(m,*)}) \stackrel{d}{\longrightarrow} \text{Normal}(\bm{0}, \mathcal{I}({\bm{\psi}}_{(m,*)}) ^{-1} \mathcal{J}({\bm{\psi}}_{(m,*)}) \mathcal{I}({\bm{\psi}}_{(m,*)}) ^{-1})
\]
as $n \longrightarrow \infty$, where
\[
\mathcal{J}(\bm{\psi}^\prime) = \left. E \left[ \left\{\frac{\partial Q(\bm{\psi})}{\partial \bm{\psi} }\right\}  \left\{ \frac{\partial Q(\bm{\psi})}{\partial \bm{\psi} } \right\}^\top \right] \right|_{\bm{\psi} = \bm{\psi}^\prime}
\]
and
\[
\bm{I} ({\bm{\psi}}^\prime) = -\frac{1}{n} \left. \frac{\partial^2 Q({\bm{\psi}})}{\partial {\bm{\psi}} \partial {\bm{\psi}}^\top } \right|_{{\bm{\psi}} = {\bm{\psi}}^\prime} \qquad \bm{J} ({\bm{\psi}}^\prime) = \frac{1}{n} \left[ \frac{\partial Q({\bm{\psi}})}{\partial {\bm{\psi}}} \left\{ \frac{\partial Q({\bm{\psi}})}{\partial {\bm{\psi}}} \right\}^\top \right]_{{\bm{\psi}} = {\bm{\psi}}^\prime}. \qquad
\]
By standard convergence results
\[
(\hat{\bm{\psi}}_{(m)} - {\bm{\psi}}_{(m,*)})^\top \bm{I} ({\bm{\psi}}_{(m,*)}) (\hat{\bm{\psi}}_{(m)} - {\bm{\psi}}_{(m,*)}) = (\hat{\bm{\psi}}_{(m)} - {\bm{\psi}}_{(m,*)})^\top \mathcal{I}({\bm{\psi}}_{(m,*)}) (\hat{\bm{\psi}}_{(m)} - {\bm{\psi}}_{(m,*)}) + o_p(1)
\]
and by a standard result for quadratic forms
\[
E[(\hat{\bm{\psi}}_{(m)} - {\bm{\psi}}_{(m,*)})^\top \mathcal{I}({\bm{\psi}}_{(m,*)}) (\hat{\bm{\psi}}_{(m)} - {\bm{\psi}}_{(m,*)})] =
\text{tr} \left\{\mathcal{J}({\bm{\psi}}_{(m,*)}) [\mathcal{I}({\bm{\psi}}_{(m,*)})]^{-1}  \right\}
+ \text{o}(1).
\]
Under standard regularity conditions on $Q(.)$, we have that $\hat{\bm{\psi}}_{(m)} \stackrel{p}{\longrightarrow} {\bm{\psi}}_{(m,*)}$ as $n \longrightarrow \infty$, and hence for large $n$ $E[-2Q(\hat{\bm{\psi}}_{(m)})] = E[-2Q({\bm{\psi}}_{(m,*)})] + o(1)$, so
\begin{align*}
\Delta(m) & =  E[-2Q({\bm{\psi}}_{(m,*)})] + E[(\hat{\bm{\psi}}_{(m)} - {\bm{\psi}}_{(m,*)})^\top \mathcal{I}({\bm{\psi}}_{(m,*)}) (\hat{\bm{\psi}}_{(m)} - {\bm{\psi}}_{(m,*)})] + o(1)\\[6pt]
& \equiv E[-2Q({\bm{\psi}}_{(m,*)})] + 2 \text{tr} \left\{\mathcal{J}({\bm{\psi}}_{(m,*)}) [\mathcal{I}({\bm{\psi}}_{(m,*)})]^{-1}  \right\} + o(1).
\end{align*}
\noindent This completes the proof.

\medskip
\noindent The asymptotic variance of estimator $\hat{\bm{\psi}}_{(m)}$ may itself be estimated by
\[
\widehat{\bm{V}}(\hat{\bm{\psi}}_{(m)} ) = n \bm{I}(\hat{\bm{\psi}}_{(m)} )^{-1} \bm{J}(\hat{\bm{\psi}}_{(m)} ) \bm{I}(\hat{\bm{\psi}}_{(m)} )^{-1}.
\]

\section{Investigating the trace term}

\begin{figure}[ht]
\includegraphics[width=\textwidth]{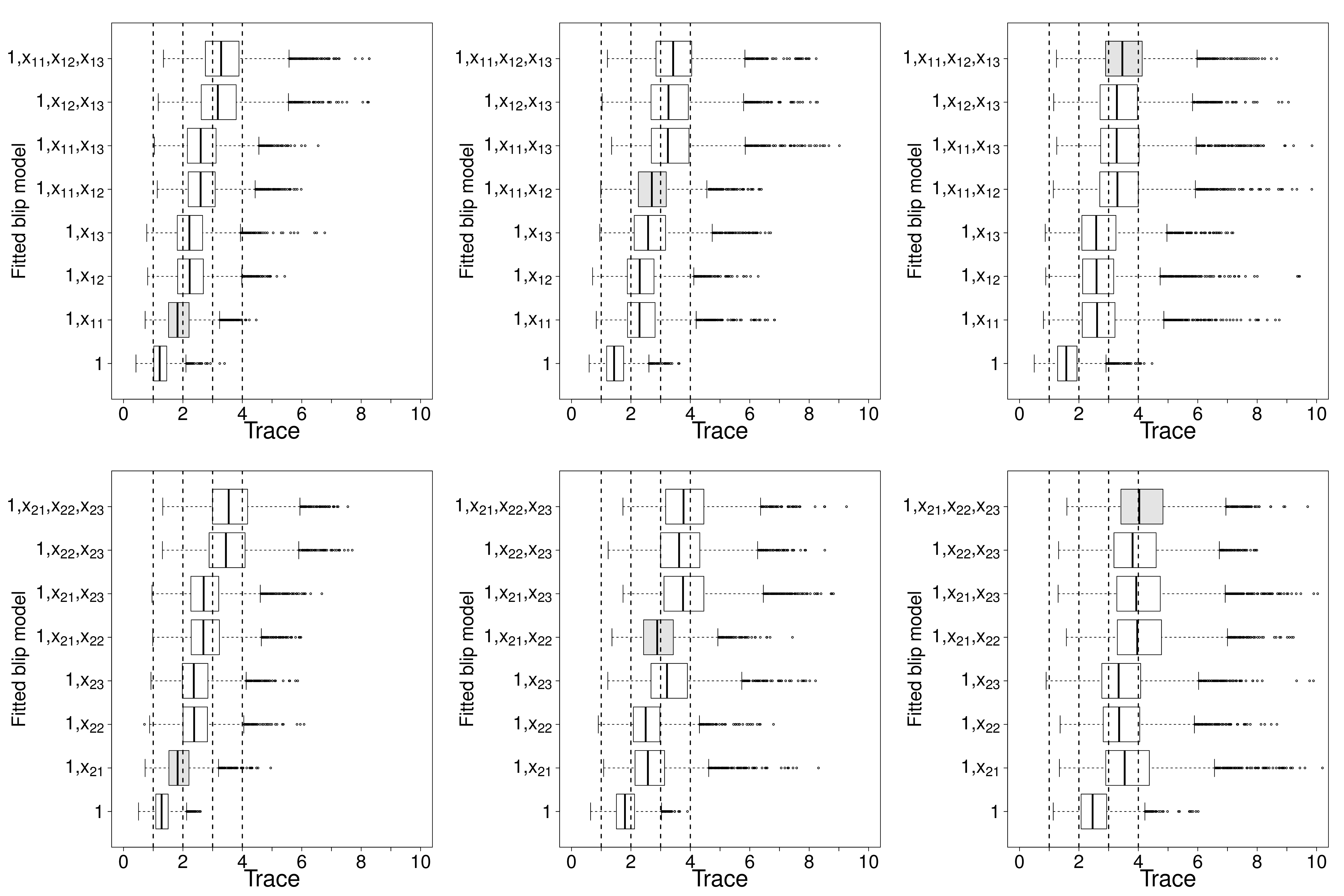}
\caption{Trace Component Term. Estimates of stage 1 (top) and stage 2 (bottom) trace term $K$ from 1,000 simulations with $n = 100$ for different candidate models (y-axis). Plots correspond to true blip models containing an intercept term along with stage $j$ covariates $x_{j1}$ (left), $x_{j1},x_{j2}$ (middle), $x_{j1},x_{j2},x_{j3}$ (right). Gray boxes correspond to simulations where the true blip model was fit.}\label{fig:figure1}
\end{figure}

We examine in simulation whether the trace term is a good approximation to the true dimension of the underlying model when that model is fitted, as described in section 5.3 of the main paper.  Figure \ref{fig:figure1} displays estimates of stage 1 (top) and stage 2 (bottom) trace term $K$ from 1,000 simulations with $n = 100$ for different candidate models (y-axis). The panels correspond to true blip models containing an intercept term along with stage $j$ covariates $x_{j1}$ (left), $x_{j1},x_{j2}$ (middle), $x_{j1},x_{j2},x_{j3}$ (right), and gray boxes correspond to simulations where the true blip model was fit.   In expectation, the trace term matches the dimension of the data generating model even in this relatively small sample setting.

\clearpage

\section{Simulation results}

\subsection{Discrete Outcome}

Next, we demonstrate the use of QIC$_G$ in the DTR framework for a discrete outcome case. We generate data as follows:

\begin{itemize}
\item stage 1 patient information: $X_{11} \sim N(1,1)$, $X_{12} \sim N(-1,1)$, $X_{13} \sim N(1,1)$;
\item stage 1 treatment: $a_1 \in \{0,1\}$, ${P}(A_1 = 1 | \bm{h}_1) = \text{expit}(x_{11})$;
\item stage 2 patient information: $X_{21} \sim N(a_1,1)$, $X_{22} \sim N(-1,1)$, $X_{23} \sim N(1,1)$;
\item stage 2 treatment: $a_2 \in \{0,1\}$, ${P}(A_2 = 1 | \bm{h}_2) = \text{expit}(x_{21})$;
\item stage $j$ blip: $\gamma_j(a_j,\bm{h}_j) = a_j(0.5 + \psi_{j1} x_{j1} + \psi_{j2} x_{j2} + \psi_{j3} x_{j3})$
\subitem such that $a_j^{opt} = 1$ if $0.5 + \psi_{j1} x_{j1} + \psi_{j2} x_{j2} + \psi_{j3} x_{j3} > 0$ and 0 otherwise;
\item outcome: $P(Y = k) = \lambda^k e^{-k}/k!$,
\subitem with $\lambda = \exp\left[\beta_0 - \sum_{j=1}^2[\gamma_j(a_j^{opt},\bm{h}_j) - \gamma_j(a_j,\bm{h}_j)]\right]$,
\end{itemize}

\noindent where we vary $\beta_0$ so that for the various $\psi_{jk}$ we consider, $P(Y = 0) = 0.1$. We set the blip parameters to $(\psi_{j1},\psi_{j2},\psi_{j3}) = (0.5,0,0)$, $(0.5,0.5,0)$ or $(0.5,0.5,0.5)$ giving a range of models including one, two, or all three variables at each stage.

Given the computational requirements for the IRLS algorithm, our simulations are somewhat more limited. We restrict ourselves to sample sizes of $n = 200$, and only consider the four blip models containing no covariates, $x_{j1}$ only, $x_{j1}$ and $x_{j2}$ only, or all three covariates, and choose whichever model resulted in the lowest $\text{QIC}_G$. As with our simulations in the main paper (Table 1), we correctly specify our treatment model, and mis-specify the treatment-free model, supposing it is linear in $x_{11}$ at stage 1 and linear in $x_{21}$ at stage 2. Results are summarized in Table \ref{tab:poisQIC}, where the stage 1 results are based on analyses where the stage 2 blip models were correctly specified. The IRLS algorithm was implemented with an iteration limit of 1,000 and a tolerance limit (between successive log-mean function estimates) of 0.001.

The presented results summarize over the simulation runs where all four IRLS algorithms converged. This corresponds to all 1,000 simulated datasets for the simplest models (with only one covariate in the true blip model at each stage), 996 (stage 2) and 994 (stage 1) datasets when two covariates were included, and 904 (stage 2) and 863 (stage 1) datasets where all three were included. If we instead presume model selection is based on the lowest QIC only for those candidate blip models where IRLS converged (as might occur in a real-life analysis), results were near-identical with the slight exception of model selection for the most complex blip model at both stages. In this case, the correct model selection rate drops from 0.921 to 0.833 for stage 2, and from 0.919 to 0.877 for stage 1, with the largest candidate model most likely to result in failure.

\begin{table}[ht]
\centering
\caption{Model selection ($n = 200$) for discrete outcome case. Entries are selection rates for the model indicated by the column heading. Bold indicates correct model selection rate.} \label{tab:poisQIC}
\begin{tabular}{lrrrr}
  \hline\hline
True & Intercept & $x_{j1}$ & $x_{j1},x_{j2}$ & $x_{j1},x_{j2},x_{j3}$\\
  \hline
$x_{11}$ & 0.017 & \textBF{0.663} & 0.169 & 0.151 \\
$x_{11},x_{12}$ & 0.001 & 0.004 & \textBF{0.799} & 0.196 \\
$x_{11},x_{12},x_{13}$ & 0.001 & 0.002 & 0.078 & \textBF{0.919} \\
$x_{21}$ & 0.005 & \textBF{0.617} & 0.189 & 0.189 \\
$x_{21},x_{22}$ & 0.001 & 0.006 & \textBF{0.836} & 0.157 \\
$x_{21},x_{22},x_{23}$ & 0.000 & 0.002 & 0.076 & \textBF{0.921} \\\hline
\end{tabular}
\end{table}

\subsection{Continuous Outcome}

Additional simulation results referred to in the main paper. Table \ref{tab:psisize} contains results for varying effect sizes, Table \ref{tab:corr} contains results for varying correlation strength between covariates, and Table \ref{tab:nonagg} contains non-aggregated results. Tables \ref{tab:normal1}-\ref{tab:normal5} contain analogous results to Tables 1-2 in the main paper and Tables \ref{tab:psisize}-\ref{tab:nonagg} in this appendix, but with standard normal errors in the generation of the outcome $Y$, rather than log-normal errors. In all tables bold indicates the most successful method for each scenario.

\begin{table}[ht]
\centering
\caption{Model selection ($n = 100$) with true blip parameters of 1, 0.5, or 0.1, under the setup described in section 3.2. (F) and (B) denote forwards and backwards selection, respectively. Bold indicates most successful approach for each setup.} \label{tab:psisize}
\begin{tabular}{llcccc}
  \hline\hline
$\psi_{jp}$ & Model & QIC$_G$ (F) & QIC$_G$ (B) & Wald (F) & Wald (B) \\
  \hline
1 & $x_{11}$ & 0.437 & 0.423 &\textBF{0.479} & 0.471 \\
   & $x_{11},x_{12}$ & 0.373 & \textBF{0.380} & 0.284 & 0.318 \\
   & $x_{11},x_{12},x_{13}$ & 0.372 & \textBF{0.419} & 0.181 & 0.254 \\
   & $x_{21}$ & 0.372 & 0.366 & \textBF{0.422} & 0.418 \\
   & $x_{21},x_{22}$ & 0.353 & \textBF{0.365} & 0.263 & 0.285 \\
   & $x_{21},x_{22},x_{23}$ & 0.357 & \textBF{0.381} & 0.150 & 0.205 \\
  0.5 & $x_{11}$ & \textBF{0.267} & 0.259 & 0.215 & 0.208 \\
   & $x_{11},x_{12}$ & 0.160 & \textBF{0.169} & 0.071 & 0.086 \\
   & $x_{11},x_{12},x_{13}$ & 0.128 & \textBF{0.165} & 0.033 & 0.063 \\
   & $x_{21}$ & \textBF{0.232} & 0.229 & 0.219 & 0.214 \\
   & $x_{21},x_{22}$ & 0.162 & \textBF{0.177} & 0.077 & 0.089 \\
   & $x_{21},x_{22},x_{23}$ & 0.132 & \textBF{0.156} & 0.028 & 0.049 \\
  0.1 & $x_{11}$ & \textBF{0.122} & 0.120 & 0.073 & 0.072 \\
   & $x_{11},x_{12}$ & 0.043 & \textBF{0.048} & 0.005 & 0.012 \\
   & $x_{11},x_{12},x_{13}$ & 0.021 & \textBF{0.037} & 0.000 & 0.002 \\
   & $x_{21}$ & \textBF{0.138} & \textBF{0.138} & 0.076 & 0.073 \\
   & $x_{21},x_{22}$ & 0.053 & \textBF{0.063} & 0.016 & 0.02 \\
   & $x_{21},x_{22},x_{23}$ & 0.030 & \textBF{0.045} & 0.003 & 0.003 \\
   \hline
\end{tabular}
\end{table}

\begin{table}[ht]
\centering
\caption{Model selection ($n = 100$) with non-treatment covariates drawn from multivariate normal distributions with correlations of 0 (`none'), 0.25 (`medium') and 0.5 (`strong'), under the setup described in section 3.2. (F) and (B) denote forwards and backwards selection, respectively. Bold indicates most successful approach for each setup.}  \label{tab:corr}
\begin{tabular}{llcccc}
  \hline\hline
$n$ & Model & QIC$_G$ (F) & QIC$_G$ (B) & Wald (F) & Wald (B) \\
  \hline
None & $x_{11}$ & 0.437 & 0.423 & \textBF{0.479} & 0.471 \\
   & $x_{11},x_{12}$ & 0.373 & \textBF{0.380} & 0.284 & 0.318 \\
   & $x_{11},x_{12},x_{13}$ & 0.372 & \textBF{0.419} & 0.181 & 0.254 \\
   & $x_{21}$ & 0.372 & 0.366 & \textBF{0.422} & 0.418 \\
   & $x_{21},x_{22}$ & 0.353 & \textBF{0.365} & 0.263 & 0.285 \\
   & $x_{21},x_{22},x_{23}$ & 0.357 & \textBF{0.381} & 0.150 & 0.205 \\
  Medium & $x_{11}$ & 0.357 & 0.353 & \textBF{0.412} & 0.408 \\
   & $x_{11},x_{12}$ & 0.331 & \textBF{0.338} & 0.247 & 0.262 \\
   & $x_{11},x_{12},x_{13}$ & 0.325 & \textBF{0.346} & 0.150 & 0.173 \\
   & $x_{21}$ & 0.351 & 0.348 & \textBF{0.388} & 0.382 \\
   & $x_{21},x_{22}$ & 0.293 & \textBF{0.301} & 0.212 & 0.223 \\
   & $x_{21},x_{22},x_{23}$ & 0.289 & \textBF{0.307} & 0.129 & 0.141 \\
  Strong & $x_{11}$ & 0.351 & 0.347 & \textBF{0.375} & 0.374 \\
   & $x_{11},x_{12}$ & 0.297 & \textBF{0.299} & 0.194 & 0.196 \\
   & $x_{11},x_{12},x_{13}$ & 0.237 & \textBF{0.244} & 0.089 & 0.093 \\
   & $x_{21}$ & 0.316 & 0.313 & \textBF{0.337} & 0.335 \\
   & $x_{21},x_{22}$ & 0.234 & \textBF{0.236} & 0.151 & 0.152 \\
   & $x_{21},x_{22},x_{23}$ & 0.196 & \textBF{0.206} & 0.065 & 0.066 \\
   \hline
\end{tabular}
\end{table}

\begin{table}[ht]
\centering
\caption{Non-aggregated selection ($n = 100$), under the setup described in section 3.2. (F) and (B) denote forwards and backwards selection, respectively. Bold indicates most successful approach for each setup.} \label{tab:nonagg}
\begin{tabular}{llllll}
  \hline\hline
Stage 1 & Stage 2 & QIC$_G$ (F) & QIC$_G$ (B) & Wald (F) & Wald (B) \\
  \hline
& & \multicolumn{4}{l}{Stage 1 Selection}\\
$x_{11}$ & $x_{21}$ & 0.437 & 0.422 & \textBF{0.492} & 0.481 \\
   & $x_{21},x_{22}$ & 0.443 & 0.429 & \textBF{0.476} & 0.467 \\
   & $x_{21},x_{22},x_{23}$ & 0.432 & 0.417 & \textBF{0.469} & 0.465 \\
  $x_{11},x_{12}$ & $x_{21}$ & 0.383 & \textBF{0.388} & 0.287 & 0.328 \\
   & $x_{21},x_{22}$ & 0.367 & \textBF{0.373} & 0.288 & 0.321 \\
   & $x_{21},x_{22},x_{23}$ & 0.370 & \textBF{0.378} & 0.276 & 0.306 \\
  $x_{11},x_{12},x_{13}$ & $x_{21}$ & 0.392 & \textBF{0.434} & 0.189 & 0.274 \\
   & $x_{21},x_{22}$ & 0.366 & \textBF{0.418} & 0.183 & 0.247 \\
   & $x_{21},x_{22},x_{23}$ & 0.359 & \textBF{0.406} & 0.171 & 0.241 \\
& & \multicolumn{4}{l}{Stage 2 Selection}\\
  $x_{11}$ & $x_{21}$ & 0.373 & 0.370 & \textBF{0.426} & 0.424 \\
  $x_{11},x_{12}$ &  & 0.375 & 0.366 & \textBF{0.418} & 0.411 \\
  $x_{11},x_{12},x_{13}$ &  & 0.367 & 0.362 & \textBF{0.421} & 0.419 \\
  $x_{11}$ & $x_{21},x_{22}$ & 0.364 & \textBF{0.375} & 0.275 & 0.299 \\
  $x_{11},x_{12}$ &  & 0.351 & \textBF{0.364} & 0.255 & 0.278 \\
  $x_{11},x_{12},x_{13}$ &  & 0.345 & \textBF{0.356} & 0.258 & 0.278 \\
  $x_{11}$ & $x_{21},x_{22},x_{23}$ & 0.358 & \textBF{0.386} & 0.162 & 0.215 \\
  $x_{11},x_{12}$ &  & 0.361 & \textBF{0.386} & 0.142 & 0.202 \\
  $x_{11},x_{12},x_{13}$ &  & 0.351 & \textBF{0.371} & 0.147 & 0.198 \\
   \hline
\end{tabular}
\end{table}

\begin{table}[ht]
\centering
\caption{Model selection for a variety of sample sizes ($n$), under the setup described in section 3.2 but with normal (rather than log-normal) errors $\epsilon \sim N(0,1)$. (F) and (B) denote forwards and backwards selection, respectively. Bold indicates most successful approach for each setup.} \label{tab:normal1}
\begin{tabular}{llcccc}
  \hline\hline
$n$ & Model & QIC$_G$ (F) & QIC$_G$ (B) & Wald (F) & Wald (B) \\
  \hline
50 & $x_{11}$ & 0.377 & 0.365 & \textBF{0.482} & 0.475 \\
   & $x_{11},x_{12}$ & 0.388 & \textBF{0.399} & 0.294 & 0.341 \\
   & $x_{11},x_{12},x_{13}$ & 0.412 & \textBF{0.470} & 0.163 & 0.253 \\
   & $x_{21}$ & 0.297 & 0.286 & \textBF{0.375} & 0.364 \\
   & $x_{21},x_{22}$ & 0.291 & \textBF{0.300} & 0.205 & 0.238 \\
   & $x_{21},x_{22},x_{23}$ & 0.291 & \textBF{0.329} & 0.126 & 0.167 \\
  100 & $x_{11}$ & 0.556 & 0.543 & \textBF{0.767} & 0.763 \\
   & $x_{11},x_{12}$ & 0.643 & 0.646 & 0.644 & \textBF{0.705} \\
   & $x_{11},x_{12},x_{13}$ & 0.796 & \textBF{0.842} & 0.513 & 0.652 \\
   & $x_{21}$ & 0.489 & 0.481 & 0.637 & \textBF{0.639} \\
   & $x_{21},x_{22}$ & 0.574 & \textBF{0.582} & 0.539 & 0.571 \\
   & $x_{21},x_{22},x_{23}$ & 0.676 & \textBF{0.706} & 0.410 & 0.485 \\
  200 & $x_{11}$ & 0.624 & 0.618 & \textBF{0.856} & 0.850 \\
   & $x_{11},x_{12}$ & 0.774 & 0.774 & 0.887 & \textBF{0.908} \\
   & $x_{11},x_{12},x_{13}$ & 0.994 & \textBF{0.995} & 0.937 & 0.974 \\
   & $x_{21}$ & 0.614 & 0.608 & \textBF{0.820} & \textBF{0.820} \\
   & $x_{21},x_{22}$ & 0.764 & 0.766 & 0.857 & \textBF{0.871} \\
   & $x_{21},x_{22},x_{23}$ & 0.958 & \textBF{0.964} & 0.865 & 0.903 \\
   \hline
\end{tabular}
\end{table}

\begin{table}[ht]
\centering
\caption{Stage 1 model selection ($n = 100$) when the stage 2 model is correctly specified (`Correct'), selected by the corresponding approach (`Recommended'), or an intercept-only model (`Intercept'), under the setup described in section 3.2 but with normal (rather than log-normal) errors $\epsilon \sim N(0,1)$. (F) and (B) denote forwards and backwards selection, respectively. Bold indicates most successful approach for each setup.}  \label{tab:normal2}
\begin{tabular}{llcccc}
  \hline\hline
Selection & Model & QIC$_G$ (F) & QIC$_G$ (B) & Wald (F) & Wald (B) \\
  \hline
Correct & $x_{11}$ & 0.556 & 0.543 & \textBF{0.767} & 0.763 \\
   & $x_{11},x_{12}$ & 0.643 & 0.646 & 0.644 & \textBF{0.705} \\
   & $x_{11},x_{12},x_{13}$ & 0.796 & \textBF{0.842} & 0.513 & 0.652 \\
  Recommended & $x_{11}$ & 0.555 & 0.541 & \textBF{0.750} & 0.748 \\
   & $x_{11},x_{12}$ & 0.636 & 0.641 & 0.627 & \textBF{0.692} \\
   & $x_{11},x_{12},x_{13}$ & 0.784 & \textBF{0.831} & 0.492 & 0.629 \\
  Intercept & $x_{11}$ & 0.552 & 0.539 & \textBF{0.721} & 0.720 \\
   & $x_{11},x_{12}$ & 0.628 & 0.633 & 0.576 & \textBF{0.640} \\
   & $x_{11},x_{12},x_{13}$ & 0.730 & \textBF{0.776} & 0.419 & 0.557 \\
   \hline
\end{tabular}
\end{table}

\begin{table}[ht]
\centering
\caption{Model selection ($n = 100$) with true blip parameters of 1, 0.5, or 0.1, under the setup described in section 3.2 but with normal (rather than log-normal) errors $\epsilon \sim N(0,1)$. (F) and (B) denote forwards and backwards selection, respectively. Bold indicates most successful approach for each setup.} \label{tab:normal3}
\begin{tabular}{llcccc}
  \hline
$\psi_{jp}$ & Model & QIC$_G$ (F) & QIC$_G$ (B) & Wald (F) & Wald (B) \\
  \hline
1 & $x_{11}$ & 0.556 & 0.543 & \textBF{0.767} & 0.763 \\
   & $x_{11},x_{12}$ & 0.643 & 0.646 & 0.644 & \textBF{0.705} \\
   & $x_{11},x_{12},x_{13}$ & 0.796 & \textBF{0.842} & 0.513 & 0.652 \\
   & $x_{21}$ & 0.489 & 0.481 & 0.637 & \textBF{0.639} \\
   & $x_{21},x_{22}$ & 0.574 & \textBF{0.582} & 0.539 & 0.571 \\
   & $x_{21},x_{22},x_{23}$ & 0.676 & \textBF{0.706} & 0.410 & 0.485 \\
  0.5 & $x_{11}$ & \textBF{0.387} & 0.379 & 0.378 & 0.377 \\
   & $x_{11},x_{12}$ & 0.308 & \textBF{0.318} & 0.164 & 0.201 \\
   & $x_{11},x_{12},x_{13}$ & 0.265 & \textBF{0.322} & 0.071 & 0.127 \\
   & $x_{21}$ & 0.297 & 0.293 & \textBF{0.310} & 0.305 \\
   & $x_{21},x_{22}$ & 0.273 & \textBF{0.280} & 0.142 & 0.158 \\
   & $x_{21},x_{22},x_{23}$ & 0.217 & \textBF{0.247} & 0.060 & 0.088 \\
  0.1 & $x_{11}$ & 0.130 & \textBF{0.127} & 0.075 & 0.076 \\
   & $x_{11},x_{12}$ & 0.039 & \textBF{0.044} & 0.006 & 0.012 \\
   & $x_{11},x_{12},x_{13}$ & 0.014 & \textBF{0.028} & 0.002 & 0.006 \\
   & $x_{21}$ & \textBF{0.132} & 0.131 & 0.104 & 0.105 \\
   & $x_{21},x_{22}$ & 0.058 & \textBF{0.064} & 0.022 & 0.024 \\
   & $x_{21},x_{22},x_{23}$ & 0.027 & \textBF{0.032} & 0.000 & 0.003 \\
   \hline
\end{tabular}
\end{table}

\begin{table}[ht]
\centering
\caption{Model selection ($n = 100$) with non-treatment covariates drawn from multivariate normal distributions with correlations of 0 (`none'), 0.25 (`medium') and 0.5 (`strong'), under the setup described in section 3.2 but with normal (rather than log-normal) errors $\epsilon \sim N(0,1)$. (F) and (B) denote forwards and backwards selection, respectively. Bold indicates most successful approach for each setup.} \label{tab:normal4}
\begin{tabular}{llcccc}
  \hline
Correlation & Model & QIC$_G$ (F) & QIC$_G$ (B) & Wald (F) & Wald (B) \\
  \hline
None & $x_{11}$ & 0.556 & 0.543 & \textBF{0.767} & 0.763 \\
   & $x_{11},x_{12}$ & 0.643 & 0.646 & 0.644 & \textBF{0.705} \\
   & $x_{11},x_{12},x_{13}$ & 0.796 & \textBF{0.842} & 0.513 & 0.652 \\
   & $x_{21}$ & 0.489 & 0.481 & 0.637 & \textBF{0.639} \\
   & $x_{21},x_{22}$ & 0.574 & \textBF{0.582} & 0.539 & 0.571 \\
   & $x_{21},x_{22},x_{23}$ & 0.676 & \textBF{0.706} & 0.410 & 0.485 \\
  Medium & $x_{11}$ & 0.510 & 0.502 & \textBF{0.666} & 0.661 \\
   & $x_{11},x_{12}$ & 0.588 & \textBF{0.594} & 0.550 & 0.570 \\
   & $x_{11},x_{12},x_{13}$ & 0.714 & \textBF{0.732} & 0.451 & 0.493 \\
   & $x_{21}$ & 0.468 & 0.460 & \textBF{0.585} & 0.584 \\
   & $x_{21},x_{22}$ & 0.509 & \textBF{0.517} & 0.443 & 0.459 \\
   & $x_{21},x_{22},x_{23}$ & 0.578 & \textBF{0.597} & 0.324 & 0.363 \\
  Strong & $x_{11}$ & 0.470 & 0.466 & 0.586 & \textBF{0.587} \\
   & $x_{11},x_{12}$ & 0.525 & \textBF{0.526} & 0.445 & 0.449 \\
   & $x_{11},x_{12},x_{13}$ & 0.563 & \textBF{0.574} & 0.282 & 0.293 \\
   & $x_{21}$ & 0.407 & 0.402 & \textBF{0.494} & 0.488 \\
   & $x_{21},x_{22}$ & 0.410 & \textBF{0.413} & 0.316 & 0.322 \\
   & $x_{21},x_{22},x_{23}$ & 0.402 & \textBF{0.410} & 0.190 & 0.178 \\
   \hline
\end{tabular}
\end{table}

\begin{table}[ht]
\centering
\caption{Non-aggregated selection ($n = 100$), under the setup described in section 3.2 but with normal (rather than log-normal) errors $\epsilon \sim N(0,1)$. (F) and (B) denote forwards and backwards selection, respectively. Bold indicates most successful approach for each setup.} \label{tab:normal5}
\begin{tabular}{llcccc}
  \hline
Stage 1 & Stage 2 & QIC$_G$ (F) & QIC$_G$ (B) & Wald (F) & Wald (B) \\
  \hline
& & \multicolumn{4}{l}{Stage 1 Selection}\\

$x_{11}$ & $x_{21}$ & 0.551 & 0.535 & \textBF{0.775} & 0.773 \\
   & $x_{21},x_{22}$ & 0.561 & 0.546 & \textBF{0.766} & 0.764 \\
   & $x_{21},x_{22},x_{23}$ & 0.557 & 0.547 & \textBF{0.762} & 0.756 \\
  $x_{11},x_{12}$ & $x_{21}$ & 0.645 & 0.647 & 0.650 & \textBF{0.712} \\
   & $x_{21},x_{22}$ & 0.637 & 0.641 & 0.634 & \textBF{0.699} \\
   & $x_{21},x_{22},x_{23}$ & 0.646 & 0.649 & 0.651 & \textBF{0.706} \\
  $x_{11},x_{12},x_{13}$ & $x_{21}$ & 0.799 & \textBF{0.839} & 0.525 & 0.663 \\
   & $x_{21},x_{22}$ & 0.794 & \textBF{0.842} & 0.507 & 0.644 \\
   & $x_{21},x_{22},x_{23}$ & 0.794 & \textBF{0.846} & 0.511 & 0.653 \\
   &  &  &  &  &  \\
& & \multicolumn{4}{l}{Stage 2 Selection}\\
  $x_{11}$ & $x_{21}$ & 0.498 & 0.490 & 0.642 & \textBF{0.643} \\
  $x_{11},x_{12}$ &  & 0.491 & 0.481 & \textBF{0.646} & 0.643 \\
  $x_{11},x_{12},x_{13}$ &  & 0.477 & 0.473 & 0.623 & \textBF{0.631} \\
  $x_{11}$ & $x_{21},x_{22}$ & 0.591 & \textBF{0.600} & 0.558 & 0.589 \\
  $x_{11},x_{12}$ &  & 0.582 & \textBF{0.589} & 0.541 & 0.576 \\
  $x_{11},x_{12},x_{13}$ &  & 0.548 & \textBF{0.556} & 0.519 & 0.548 \\
  $x_{11}$ & $x_{21},x_{22},x_{23}$ & 0.702 & \textBF{0.729} & 0.443 & 0.512 \\
  $x_{11},x_{12}$ &  & 0.667 & \textBF{0.700} & 0.408 & 0.484 \\
  $x_{11},x_{12},x_{13}$ &  & 0.658 & \textBF{0.688} & 0.380 & 0.460 \\
   \hline
\end{tabular}
\end{table}

\bibliographystyle{jasa2}
\bibliography{longtitles,referencesforabb}{}

\end{document}